\documentclass[%
 aip,
 amsmath,amssymb,
 reprint,%
]{revtex4-1}
\usepackage{graphicx}
\usepackage{subfigure}
\usepackage{dcolumn}
\usepackage{bm}
\usepackage{color}
\usepackage[T1]{fontenc}
\usepackage{mathptmx}
\usepackage{multirow}
\usepackage{array}
\usepackage{float}
\usepackage{etoolbox}
\usepackage{booktabs}
\usepackage{url}
\usepackage{CJKutf8}
\usepackage{pifont}
\usepackage{epstopdf, epsfig}
\usepackage{graphicx}
\usepackage{longtable}
\usepackage[
    colorlinks=true,
    linkcolor=blue,
    urlcolor=blue,
    citecolor=blue
]{hyperref}

\usepackage{titlesec}
\titleclass{\subsubsubsection}{straight}[\subsection]
\newcounter{subsubsubsection}[subsubsection]
\renewcommand\thesubsubsubsection{\thesubsubsection.\alph{subsubsubsection}}
\titleformat{\subsubsubsection}
 {\normalfont\normalsize\bfseries}{\thesubsubsubsection}{1em}{}
\titlespacing*{\subsubsubsection}
{0pt}{3.25ex plus 1ex minus .2ex}{1.5ex plus .2ex}

\begin{document}
\preprint{AIP/123-QED}


\title{Investigating the Effect of Relaxation Time on Richtmyer--Meshkov Instability under Reshock Impact: A Two-Component Discrete Boltzmann Method Study}

\author{Lingyan Lian \begin{CJK*}{UTF8}{gbsn} (连玲艳) \end{CJK*}}
 \affiliation{School of Mathematics and Statistics, Key Laboratory of Analytical Mathematics and Applications (Ministry of Education), Fujian Key Laboratory of Analytical Mathematics and Applications (FJKLAMA), Center for Applied Mathematics of Fujian Province (FJNU), Fujian Normal University, 350117 Fuzhou, China}
\author{Chuandong Lin \begin{CJK*}{UTF8}{gbsn} (林传栋) \end{CJK*}}
 \thanks{Corresponding author: linchd3@mail.sysu.edu.cn}
  \affiliation{Sino-French Institute of Nuclear Engineering and Technology, Sun Yat-sen University, Zhuhai 519082, China}
  \affiliation{Key Laboratory for Thermal Science and Power Engineering of Ministry of Education, Department of Energy and Power Engineering, Tsinghua University, Beijing 100084, China}
\author{Demei Li \begin{CJK*}{UTF8}{gbsn} (李德梅) \end{CJK*}}
  \affiliation{School of Mathematics and Statistics, Key Laboratory of Analytical Mathematics and Applications (Ministry of Education), Fujian Key Laboratory of Analytical Mathematics and Applications (FJKLAMA), Center for Applied Mathematics of Fujian Province (FJNU), Fujian Normal University, 350117 Fuzhou, China}
\author{Huilin Lai \begin{CJK*}{UTF8}{gbsn} (赖惠林)\end{CJK*}}
 \thanks{Corresponding author: hllai@fjnu.edu.cn}
\affiliation{School of Mathematics and Statistics, Key Laboratory of Analytical Mathematics and Applications (Ministry of Education), Fujian Key Laboratory of Analytical Mathematics and Applications (FJKLAMA), Center for Applied Mathematics of Fujian Province (FJNU), Fujian Normal University, 350117 Fuzhou, China}

\date{\today}

\begin{abstract}
The Richtmyer--Meshkov (RM) instability plays an important role in various natural and engineering fields, such as inertial confinement fusion. In this work, the effect of relaxation time on the RM instability under reshock impact is investigated by using a two-component discrete Boltzmann method. The hydrodynamic and thermodynamic characteristics of the fluid system are comprehensively analyzed from the perspectives of the density gradient, vorticity, kinetic energy, mixing degree, mixing width, and non-equilibrium intensity. Simulation results indicate that for larger relaxation time, the diffusion and dissipation are enhanced, the physical gradients decrease, and the growth of the interface is suppressed. Furthermore, the non-equilibrium manifestations show complex patterns, driven by the competitive physical mechanisms of the diffusion, dissipation, shock wave, rarefaction wave, transverse wave, and fluid instabilities. These findings provide valuable insights into the fundamental mechanism of compressible fluid flows.
\end{abstract}

\maketitle

\section{Introduction}
The Richtmyer--Meshkov (RM) instability is a critical phenomenon in fluid dynamics that arises when a shock wave propagates through the interface between two fluids of different densities. The misalignment between the pressure and density gradients generates baroclinic vorticity, which amplifies disturbances at the interface. This amplification leads to the formation of complex flow structures, such as bubbles and spikes, which evolve over time and ultimately promote the mixing of the two fluids. The RM instability is widely observed in both natural phenomena and engineering applications, including weapon implosion \cite{schill2024suppression}, inertial confinement fusion (ICF) \cite{zhou2025instabilities}, supernova explosions \cite{abarzhi2024perspective}, and astrophysics \cite{zhou2024hydrodynamic}. In ICF, the RM instability plays an important role in influencing fuel compression efficiency and ignition symmetry. As shock waves propagate through the multi-layer interfaces within the fuel chamber, the RM instability triggers the growth of interface disturbances, leading to turbulent mixing, fuel contamination, and shock wave propagation asymmetry. These effects ultimately reduce fusion efficiency and jeopardize the success of ignition \cite{mostert2015effects}. In scramjets, the RM instability accelerates the growth of interface disturbances, which promotes turbulent mixing and enhances the fuel-air mixing efficiency, ultimately improving combustion performance. However, excessive RM instability can lead to unstable combustion, which may adversely affect engine operation \cite{attal2015numerical}. Therefore, a deeper understanding of the underlying physical mechanisms of RM instability is essential for optimizing engineering applications.

The RM instability was first theoretically proposed by Richtmyer in 1960, providing a foundational framework for understanding the growth mechanism of density interface instability under the influence of shock waves \cite{1960Richtmyer}. This theoretical framework was subsequently validated by Meshkov in 1969 through experimental shock tube studies, offering pivotal experimental evidence that supported Richtmyer’s theoretical predictions \cite{meshkov1969instability}. Since then, the field has attracted considerable attention, with numerous researchers dedicating efforts to exploring the mechanisms and evolving characteristics of the RM instability. However, most studies have predominantly focused on interface instability induced by a single shock wave \cite{guo2022shock,wang2023high,latini2024analysis}. In contrast, in complex flow scenarios, shock waves frequently interact with density interfaces multiple times, with reflected shock waves playing a pivotal role in these interactions. Compared to single shock waves, the introduction of reflected shocks can significantly alter the dynamic characteristics of interface instability, such as accelerating the growth of the mixing layer and intensifying turbulence, thus drawing significant research interest. On the theoretical front, in 1989, Brouillette \textit{et al}. introduced a generalized Richtmyer model to explain the cumulative effect of multiple shocks and to elucidate the nonlinear growth of perturbations through both theoretical and experimental analyses \cite{brouillette1989growth}. In 2001, Charakhch'yan \textit{et al}. performed numerical simulations of two-dimensional RM instability to investigate the influence of reflected shocks on the growth of interface perturbations, proposing a perturbation amplitude model applicable under reflected shock conditions \cite{charakhch2001reshocking}. In 1989, Mikaelian \textit{et al}. introduced a reflection shock model to describe the impact of reflected shocks on interface disturbances \cite{mikaelian1989turbulent}, although this model was initially limited to the early stages of reflection shock interactions. To overcome this limitation, Mikaelian \textit{et al}. refined the model in 2011, broadening its applicability \cite{mikaelian2011extended}. Further enhancing the model's ability to accurately simulate complex phenomena during the reflection shock process, Mikaelian optimized it again in 2015 \cite{mikaelian2015testing}.  Although existing theoretical methods have made progress in studying RM instability under reflected shock conditions, they still struggle to fully capture its complex multi-scale effects and nonlinear behavior.

In experimental studies, the RM instability is roughly divided into several key factors, including interface shape \cite{zhang2023richtmyer}, interface disturbance \cite{hosseini2005experimental,jacobs2013experiments,guo2024richtmyer}, incident shock wave type \cite{zhai2018review,chen2024numerical} and so on. These factors have a significant influence on the instability caused by reflected shock waves and the evolution of the mixed region. In 2005, Hosseini \textit{et al}. studied the RM instability induced by cylindrical shocks via shock tube experiments, showing that initial perturbation shape and amplitude affect mixing evolution, with shock convergence intensifying instability and turbulence \cite{hosseini2005experimental}. In 2009, Leinov \textit{et al}. studied fluid mixing induced by the RM instability under reshock conditions using shock tube experiments and simulations. They found that reshock strength significantly affects mixing evolution and proposed a model linking bubble competition to the linear growth rate \cite{leinov2009experimental}. In 2012, Si \textit{et al}. investigated the evolution of spherical gas interfaces under reshock waves through shock tube experiments. It was shown that the reshock induces further instability of the interface, leading to rapid growth of the mixing zone and enhancing the turbulent mixing characteristics of the interface \cite{si2012experimental}. In 2021, Ding \textit{et al}. used a semi-annular shock tube to study the effects of layer thickness and phase difference on instability evolution. They observed that greater layer thickness accelerates outer interface development, while phase difference critically influences inner interface growth \cite{ding2021convergent}. In 2023, Zhang \textit{et al}. examined the RM instability under reshock via shock tube experiments. Their findings revealed that reshock amplifies interface instability, with complex initial disturbances enhancing turbulent mixing and increasing the mixing layer growth rate \cite{zhang2023richtmyer}.  In 2024, Ferguson \textit{et al}. discovered that shorter intervals between shock and reshock accelerate growth and mixing, while longer intervals reduce reshock effects \cite{ferguson2024influence}. Although experimental studies have made significant contributions to understanding RM instability under reflected shock conditions, the limitations of experimental setups still make it difficult to fully reveal the complex physical mechanisms involved.

In addition to theoretical analysis and experimental research, numerical simulations have also played a crucial role in the study of the RM instability, such as finite volume method  \cite{chen2023numerical,chen2024numerical}, large eddy simulation \cite{wang2010large,lombardini2011atwood,bin2021new,sadler2024simulations} and so on. With the advancement of high-precision computational techniques, the discrete Boltzmann method (DBM) has shown unique advantages in studying the RM instability, emerging as a promising tool for investigating complex flow phenomena \cite{xu2024advances,gan2018discrete}. In recent years, DBM has become an important tool for investigating complex fluid systems, such as compressible reacting flows \cite{lin2017multi,lin2020kinetic,ji2022three,su2023unsteady,lin2024discrete}, multiphase flows \cite{gan2015discrete,zhang2019entropy,gan2022discrete,sun2024droplet}, Rayleigh--Taylor (RT) instability \cite{ye2020knudsen,chen2021specific,li2022rayleigh,lai2024investigation,chen2024surface}, and Kelvin--Helmholtz (KH) instability \cite{gan2019nonequilibrium,li2022influence,li2024kinetic,xu2025influence}. DBM originates from the lattice Boltzmann method (LBM) and is often regarded as a variant of LBM \cite{osborn1995lattice,swift1995lattice,liang2014phase,liang2016lattice,silva2025analysis}. However, the traditional LBM has primarily been used to solve fluid dynamics equations and has seen limited application in the kinetic modeling of non-equilibrium systems. As a mesoscopic method, DBM bridges the gap between macroscopic and microscopic approaches. Its kinetic framework focuses on moments closely related to the evolution of complex systems, enabling it to capture key non-equilibrium information. This capability makes DBM particularly effective in multi-physics coupling and complex flow studies, offering a novel perspective and method for exploring non-equilibrium effects. In 2012, Xu \textit{et al}. proposed to describe the complex thermodynamic non-equilibrium (TNE) behavior in systems by means of non-conserved moments of $(f-f^{eq})$ \cite{xu2012lattice}. In 2014, Lin \textit{et al}. proposed a polar coordinate DBM to simulate shock wave induced the RM instability, focusing on non-equilibrium properties near mechanical and material interfaces \cite{lin2014polar}. In 2016, Lin \textit{et al}. proposed the two-component DBM to simulate shock wave induced the RM instability \cite{lin2016double}. In 2018, Chen \textit{et al}. explored the interaction of RM and RT instabilities through the multi-relaxation discrete Boltzmann model, revealing the effects of gravity and Mach number on the synergy and competition of instabilities \cite{chen2018collaboration}. In 2023, Shan \textit{et al}. used the DBM to analyze the non-equilibrium kinetics effects of the RM instability and reshock wave process, revealing that shock waves cause significant deviations from thermodynamic equilibrium at the material interface, with complex kinetic effects observed through two sets of TNE quantities, which highlight distinct entropy production behaviors in different fluids \cite{shan2023nonequilibrium}. In the same year, Yang \textit{et al}. used the DBM to investigate the influence of density ratio on the RM instability and non-equilibrium effect in the reshock process. The result shows that the density ratio significantly affects the evolution characteristics and non-equilibrium strength of the mixed layer \cite{yang2023influence}. In 2024, Song \textit{et al}. proposed a DBM for plasma kinetics to investigate non-equilibrium behaviors in RM instability problems. The study explores entropy production rates and TNE effects, emphasizing the role of magnetic fields in influencing interface inversion \cite{song2024plasma}. DBM provides a powerful tool for studying RM instability, effectively capturing complex nonequilibrium effects and offering new perspectives and methods for further understanding complex phenomena in fluid dynamics.

To date, many scholars have numerically studied the RM instability under various conditions, including the effects of diffusion and viscosity \cite{weber2014inhibition,pereira2021molecular}, further deepening our understanding of its complex dynamics and providing valuable insights into the role of these factors in influencing instability behavior. In 1993, Mikaelian \textit{et al}. investigated the impact of viscosity on the RM instability and found that viscosity dissipates vorticity and energy, thereby suppressing nonlinear interface growth and influencing the expansion of the mixing layer and turbulence intensity \cite{mikaelian1993effect}. In 2009, Sohn \textit{et al}. examined the effects of surface tension and viscosity on the RM instability, revealing that surface tension stabilizes the interface by inhibiting disturbance growth, while viscosity dissipates turbulent energy, slowing nonlinear growth and significantly affecting mixing layer expansion and turbulence characteristics \cite{sohn2009effects}. In 2017, Walchli \textit{et al}. used numerical simulations to explore the effect of Reynolds number (\textit{Re}) on the nonlinear growth rate of the RM instability by varying fluid viscosity. They found that lower Reynolds numbers significantly extended the time to reach nonlinear saturation, with the most pronounced effects observed for \textit{Re} $< 256$ \cite{walchli2017reynolds}. In 2019, Wong \textit{et al}. employed high-resolution Navier--Stokes (NS) simulations to study the RM instability under re-shock conditions, discovering that viscosity and diffusion suppress small-scale turbulence, slow mixing layer expansion, and reduce turbulence intensity \cite{wong2019high}. In 2020, Liu \textit{et al}. investigated the circulation deposition process during the RM instability both theoretically and numerically. They quantified the impact of viscosity on circulation deposition in the RM instability and revealed the crucial role of viscosity in controlling turbulence mixing and interface stability \cite{liu2020contribution}. The same year, Sun proposed a unified model to study the effects of elasticity, viscosity, and magnetic fields on linear the RM instability. The study explored how these three factors individually or jointly influence the growth rate and interface evolution of the RM instability \cite{sun2020unified}. The results showed that viscosity primarily affects turbulence and mixing processes through dissipation mechanisms.

However, there is little study on the impact of the relaxation time on the RM instability. The relaxation time not only characterizes the viscous dissipation properties of the fluid but also serves as an indicator of non-equilibrium effects on fluid dynamics, which can substantially influence the behavior of shock-interface interactions. Therefore, studying the relaxation time contributes to further revealing the non-equilibrium characteristics during the interface evolution of RM instability, reflecting the non-equilibrium features of shock wave-material interface interactions, deepening the understanding of interface instability in complex flows, and providing theoretical references for relevant engineering applications. In this work, the DBM is employed to perform high-resolution numerical simulations of the RM instability evolution during the reflected shock process, with particular emphasis on systematically examining relaxation time on both hydrodynamic non-equilibrium (HNE) and TNE effects. The structure of this paper is organized as follows. In Sec. \ref{SecII}, the DBM is introduced briefly. In Sec. \ref{SecIII}, the RM instability is simulated and analyzed under varying relaxation times. Finally, the conclusions are summarized in \ref{SecIV}.

\section{Discrete Boltzmann method}\label{SecII}

In this work, the governing equation of the DBM is based on the two-component Bhatnagar--Gross--Krook discrete Boltzmann equation \cite{sofonea2001bgk,lin2016double}, which can be expressed as:
\begin{equation}\label{DBM}
	\frac{\partial f^{\sigma}_{i}}{\partial t}+\pmb{v}_{i}\cdot \frac{\partial f^{\sigma}_{i}}{\partial \pmb{r}}=-\frac{1}{\tau ^{\sigma}}\left(f^{\sigma}_{i}-f^{\sigma eq}_{i}\right)\text{,}
\end{equation}
where $t$ stands for time, $\pmb{r}$ the space coordinate, $\pmb{v}_{i}$ the discrete velocity, $f^{\sigma}_{i}$ $(f^{\sigma eq}_{i})$ the discrete (local equilibrium) distribution function. The superscript $\sigma=A$ or $B$ denotes the component, and the subscript $i\,(= 1, 2, \cdots, N)$ represents the index of the discrete velocity set, where $N$ is the total number of discrete velocities. It should be noted that the relaxation time should satisfy the relationship $\tau^A=\tau^B$ to ensure that local conservation of momentum \cite{sofonea2001bgk}. For this purpose, the following expression is adopted for the relaxation time:
\begin{equation}
	\tau^{\sigma}=\left(\frac{n^{A}}{\theta^{A}}+\frac{n^{B}}{\theta^{B}}\right)^{-1}\text{,}
\end{equation}
where $\theta^{A}$ and $\theta^{B}$ indicate relaxation parameters for components $A$ and $B$, respectively.

For individual components, the particle number density $n^{\sigma}$, fluid mass density $\rho^{\sigma}$ and flow velocity $\pmb{u}^{\sigma}$can be obtained by
\begin{equation}\label{e2}
	n^{\sigma}=\sum_{i}f^{\sigma}_{i} \text{,}
\end{equation}
\begin{equation}
	\rho^{\sigma}=n^{\sigma}m^{\sigma}\text{,}
\end{equation}
\begin{equation}\label{e3}
	\pmb{u}^{\sigma}=\frac{1}{n^{\sigma}}\sum_{i}f^{\sigma}_{i}\pmb{v}_{i}\text{,}
\end{equation}
where $m^{\sigma}$ denotes the particles mass. In the total system, the number density $n$, mass density $\rho$, and flow velocity $\pmb{u}$  are expressed as follows
\begin{equation}\label{e6}
	n=\sum_{\sigma}n^{\sigma}\text{,}
\end{equation}
\begin{equation}\label{e10}
	\rho=\sum_{\sigma}\rho^{\sigma}\text{,}
\end{equation}
\begin{equation}\label{e7}
	\pmb{u}=\frac{1}{\rho}\sum_{\sigma}\rho^{\sigma}\pmb{u}^{\sigma}\text{.}
\end{equation}

The internal energy density of each component and the total internal energy density are given by
\begin{equation}\label{e5}
	E^{\sigma}=\frac{1}{2}m^{\sigma}\sum_{i}f^{\sigma}_{i}\left(\left|\pmb{v}_{i}-\pmb{u}\right|^{2}+\eta^{2}_{i}\right)\text{,}
\end{equation}
\begin{equation}\label{e9}
	E=\sum_{\sigma}E^{\sigma}\text{,}
\end{equation}
where the symbol $\eta_{i}$ represents the describe the internal energy density for extra degrees of
freedom. Then the specific temperature $T^{\sigma}$ and mixing temperature $T$ can be calculated as
\begin{equation}\label{e4}
	T^{\sigma}=\frac{2E^{\sigma}}{n^{\sigma}\left(D+I\right)}\text{,}
\end{equation}
\begin{equation}\label{e8}
	T=\frac{2E}{n\left(D+I\right)}\text{,}
\end{equation}
where $D$ represents the number of dimensions, and $I$ denotes the number of extra degrees of freedom. In the current investigation, we consider
$D=2$ and $I=3$ for our numerical simulations.

Through the Chapman--Enskog (CE) expansion, the discrete Boltzmann equation is simplified to NS equation under the assumption of continuum limit (refer to Appendix \ref{A}). To this end, the discrete equilibrium distribution function $ f^{\sigma eq}_{i}$ should satisfy the following seven moment constraints \cite{watari2004possibility}:
\begin{equation}\label{e11}
	\sum_{i}f^{\sigma eq}_{i}=n^{\sigma}\text{,}
\end{equation}
\begin{equation}\label{e12}
	\sum_{i}f^{\sigma eq}_{i}\pmb{v}_{i}=n^{\sigma}\pmb{u}\text{,}
\end{equation}	
\begin{equation}\label{e13}
	\sum_{i}f^{\sigma eq}_{i}\left(\pmb{v}_{i}\cdot \pmb{v}_{i}+\eta^{2}_{i} \right)=n^{\sigma}\left[\left(D+I\right)\frac{T}{m^{\sigma}}+\pmb{u}\cdot \pmb{u}\right]\text{,}
\end{equation}	
\begin{equation}\label{e14}
	\sum_{i}f^{\sigma eq}_{i}\pmb{v}_{i}\pmb{v}_{i}=n^{\sigma}\left(\frac{T}{m^{\sigma}}\pmb{e}_{\alpha}\pmb{e}_{\beta}\delta_{\alpha \beta}+\pmb{u}\pmb{u}\right)\text{,}
\end{equation}	
\begin{equation}\label{e15}
	\sum_{i}f^{\sigma eq}_{i}\left(\pmb{v}_{i}\cdot \pmb{v}_{i}+\eta^{2}_{i} \right)\pmb{v}_{i}=n^{\sigma}\pmb{u}\left[\left(D+I+2\right)\frac{T}{m^{\sigma}}+\pmb{u}\cdot \pmb{u}\right]\text{,}
\end{equation}	
\begin{eqnarray}\label{e16}
	&\sum_{i}f^{\sigma eq}_{i}\pmb{v}_{i}\pmb{v}_{i}\pmb{v}_{i}=n^{\sigma}[\frac{T}{m^{\sigma}}(\pmb{u}_{\alpha}\pmb{e}_{\beta}\pmb{e}_{\chi}\delta_{\delta \chi}+\pmb{u}_{\beta}\pmb{e}_{\alpha}\pmb{e}_{\chi}\delta_{\alpha \chi}\nonumber\\ [8pt]
	&+\pmb{u}_{\chi}\pmb{e}_{\alpha}\pmb{e}_{\beta}\delta_{\alpha \beta })+\pmb{u}\pmb{u}\pmb{u}]\text{,}
\end{eqnarray}

\begin{eqnarray}\label{e17}
		\sum_{i}f^{\sigma eq}_{i}\left(\pmb{v}_{i}\cdot \pmb{v}_{i}+\eta^{2}_{i} \right)\pmb{v}_{i}\pmb{v}_{i}
		&=n^{\sigma}\frac{T}{m^{\sigma}}\left[\left(D+I+2\right)\frac{T}{m^{\sigma}}
		+\pmb{u}\cdot \pmb{u}\right]\pmb{e}_{\alpha}\pmb{e}_{\beta}\delta_{\alpha \beta}\nonumber\\ [8pt]
		&+n^{\sigma}\pmb{u}\pmb{u}\left[\left(D+I+4\right)\frac{T}{m^{\sigma}}\pmb{u} \cdot \pmb{u}\right]\text{,}
\end{eqnarray}
where $\pmb{e}_{\alpha}$ is the unit vector in the direction of $\alpha$, and $\delta_{\alpha \beta}$ is the Kronecker delta function. The indices $\alpha, \beta, \chi $ represent the spatial directions $x$ or $y$. Here, the local equilibrium distribution function $f^{\sigma eq}$ is adopted as follows:
\begin{equation}\label{e255}
	f^{\sigma eq}=n^{\sigma}\Big(\frac{m^\sigma}{2\pi T}\Big)^{D/2}\Big(\dfrac{m^\sigma}{2\pi IT}\Big)^{1/2}  \exp \Big(-\frac{m^\sigma\mid\pmb{v}-\pmb{u}\mid^2}{2T}-\frac{m^\sigma \eta ^{2}}{2IT}\Big).
\end{equation}
Consequently, the seven moments above can also be written as the following matrix form:
\begin{equation}\label{e18}
	\pmb{C}\cdot\pmb{f}^{\sigma eq}=\hat{\pmb{f}}^{\sigma eq}\text{,}
\end{equation}
where
\begin{equation}\label{e19}
	\pmb{f}^{\sigma eq}=\left[f^{\sigma eq}_{1},f^{\sigma eq}_{2},\cdots ,f^{\sigma eq}_{N}\right]^{\rm{T}}\text{,}
\end{equation}
\begin{equation}\label{e20}
	\hat{\pmb{f}}^{\sigma eq}=\left[\hat{f}^{\sigma eq}_{1},\hat{f}^{\sigma eq}_{2},\cdots ,\hat{f}^{\sigma eq}_{N}\right]^{\rm{T}}\text{,}
\end{equation}
and $\pmb{C}$ is a $N\times N$ matrix, with its elements determined by a discrete velocity mode. According to Eq. (\ref{e18}), the discrete equilibrium distribution function can be obtained by
\begin{equation}\label{e22}
	\pmb{f}^{\sigma eq}=\pmb{C}^{-1}\hat{\pmb{f}}^{\sigma eq}\text{,}
\end{equation}
where the superscripts ``\,$\rm{T}$\,'' and ``\,$-1$\,'' represent the transpose and inverse of the matrix, respectively.

In the seven moment relations (\ref{e11})-(\ref{e17}), the first three correspond to conserved moments, which represent the conservation of mass, momentum, and energy, respectively. The remaining four are non-conserved moments, meaning that when $f^{\sigma}_{i}$ replaces $f^{\sigma eq}_{i}$, a discrepancy may occur between the two sides of the equations. For clarity, the distribution function $f^{\sigma}_{i}$ is decomposed into equilibrium and non-equilibrium components as follows:
\begin{equation}
	f^{\sigma}_{i}=f^{\sigma eq}_{i}+f^{\sigma neq}_{i}\text{.}
	\label{fneq}
\end{equation}
In fact, the kinetic moments of $f^{\sigma neq}_{i}$ can be used to quantify the non-equilibrium characteristics of complex flow systems, providing a more accurate description of the TNE effects. Mathematically, the non-equilibrium quantities can be defined by
\begin{equation}\label{m155}
	\pmb{\Delta}^{\sigma*}_{m}=m^{\sigma}\sum_{i}f^{\sigma neq}_{i} \underbrace{\pmb{v}^{*}_{i}\pmb{v}^{*}_{i}\cdots \pmb{v}^{*}_{i}}_{m}\text{,}
\end{equation}

\begin{equation}\label{mn01}
	\pmb{\Delta}^{\sigma*}_{m,n}=m^{\sigma}\sum_{i}f^{\sigma neq}_{i} \left(\pmb{v}^{*}_{i}\cdot \pmb{v}^{*}_{i}+\eta^{2}_{i}\right)^{\left(m-n\right)/2}\underbrace{\pmb{v}^{*}_{i}\pmb{v}^{*}_{i}\cdots \pmb{v}^{*}_{i}}_{n}\text{,}
\end{equation}
with $\pmb{v}^{*}_{i}=\pmb{v}_{i}-\pmb{u}$ denotes the peculiar velocity.

What's more, in order to better describe the global TNE of the fluid system, the following TNE quantities are defined as follows:
\begin{equation}
	|\pmb{\Delta}_{2}^{\sigma *}|=\sqrt{|\pmb{\Delta}_{2xx}^{\sigma *}|^{2}+|\pmb{\Delta}_{2xy}^{\sigma *}|^{2}+|\pmb{\Delta}_{2yy}^{\sigma *}|^{2}}\text{,}
\end{equation}
\begin{equation}
	|\pmb{\Delta}_{3,1}^{\sigma *}|=\sqrt{|\pmb{\Delta}_{3,1x}^{\sigma *}|^{2}+|\pmb{\Delta}_{3,1y}^{\sigma *}|^{2}}\text{,}
\end{equation}
\begin{equation}
	|\pmb{\Delta}_{3}^{\sigma *}|=\sqrt{|\pmb{\Delta}_{3xxx}^{\sigma *}|^{2}+|\pmb{\Delta}_{3xxy}^{\sigma *}|^{2}+|\pmb{\Delta}_{3xyy}^{\sigma *}|^{2}+|\pmb{\Delta}_{3yyy}^{\sigma *}|^{2}}\text{,}
\end{equation}
\begin{equation}
	|\pmb{\Delta}_{4,2}^{\sigma *}|=\sqrt{|\pmb{\Delta}_{4,2xx}^{\sigma *}|^{2}+|\pmb{\Delta}_{4,2xy}^{\sigma *}|^{2}+|\pmb{\Delta}_{4,2yy}^{\sigma *}|^{2}}\text{,}
\end{equation}
\begin{equation}
	|\pmb{\Delta}^{\sigma *}|=\sqrt{|\pmb{\Delta}_{2}^{\sigma *}|^{2}+|\pmb{\Delta}_{3,1}^{\sigma *}|^{2}+|\pmb{\Delta}_{3}^{\sigma *}|^{2}+|\pmb{\Delta}_{4,2}^{\sigma *}|^{2}}\text{,}
\end{equation}
where $\pmb{\Delta}_{2}^{\sigma *}$ signifies the non-organized momentum flux, which is connected to the viscosity, $\pmb{\Delta}_{3,1}^{\sigma *}$ and $\pmb{\Delta}_{3}^{\sigma *}$ denote the non-organized energy flux, which are related to the heat flux, $\pmb{\Delta}_{4,2}^{\sigma *}$  represents the flux of the non-organized energy flux, and $\pmb{\Delta}^{\sigma *}$ stands for the global TNE quantities, which describe the extent of deviation from the system's equilibrium state. The global average TNE intensity can be obtained by integrating the global TNE quantities and averaging over the computational domain
\begin{equation}\label{e25}
	\overline{D}^{\sigma^{*}}=\frac{1}{L_{x}L_{y}}\int_{0}^{L_{x}}\int_{0}^{L_{y}}\left|\pmb{\Delta}^{\sigma *}\right|\rm{d}x\rm{d}y\text{,}
\end{equation}
where $L_{x}$ and $L_{y}$ represent the length and width of the fluid system, respectively.

It should be mentioned that the TNE effects are associated with $f^{\sigma neq}_{i}$, and the HNE effects are related to the spacial variation of physical quantities, such as the physical gradients of the concentration, density, velocity, temperature and pressure. Mathematically, using the definition in Eq. (\ref{fneq}), Eq. (\ref{DBM}) can be rewritten in a more intuitive form, explicitly distinguishing between the equilibrium and non-equilibrium parts:
\begin{equation}
	\frac{{\partial{f_i}^{\sigma}}}{{\partial t}} + \pmb{v}_{i}\cdot\dfrac{{\partial {f_i}^{\sigma eq}}}{{\partial {\pmb{r}}}}=-\left(\dfrac{1}{\tau^{\sigma}}+\pmb{v}_{i}\cdot\dfrac{\partial }{{\partial {\pmb{r}}}}\right) {f_i}^{\sigma neq}\text{.}
	\label{DBM_neq}
\end{equation}
Physically, the TNE effects arise from the term on the right-hand side of Eq. (\ref{DBM_neq}), and the HNE effects originate entirely from the second term on the left-hand side of the equation. According to the CE multi-scale analysis, the spatial derivatives in Eq. (\ref{DBM_neq}) corresponds to the gradients of macroscopic physical quantities, which are directly related to the formation of HNE effects. In addition, on the NS level, the convection term includes the effects of viscosity and heat conduction, which are a coarse-grained description of THE effects. At the Euler level, $f_{i}^{\sigma neq}$ approaches zero, and the discrete Boltzmann equation can be only used to capture the HNE effects.
\begin{figure}[htbp]
	{\centering
		\includegraphics[width=0.4\textwidth]{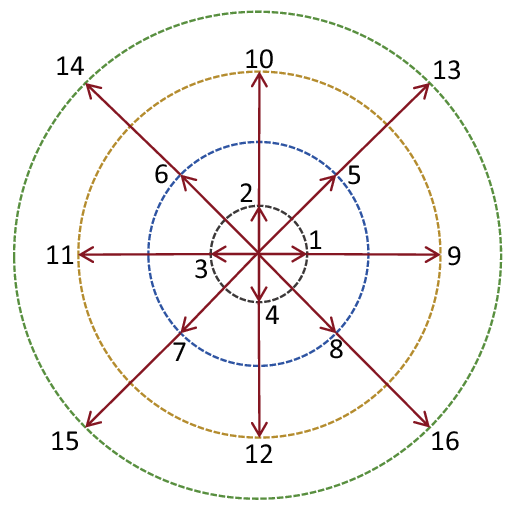}}
	\caption{\centering{Sketch of D2V16 discrete velocity model}}
	\label{Fig01}
\end{figure}

As shown in Fig. \ref{Fig01}, the D2V16 discrete velocity model is adopted as \cite{lin2019discrete}
\begin{equation}\label{e26}
	\pmb{v}_{i}=
	\left\{
	\begin{aligned}
		&v_{a}\Big[\cos\frac{\left(i-1\right)\pi}{2}\text{,}\sin\frac{\left(i-1\right)\pi}{2}\Big]\text{,}~1\leq i \leq 4\text{,}\\
		&v_{b}\Big[\cos\frac{\left(2i-1\right)\pi}{4}\text{,}\sin\frac{\left(2i-1\right)\pi}{4}\Big]\text{,}~5\leq i \leq 8\text{,}\\
		&v_{c}\Big[\cos\frac{\left(i-9\right)\pi}{2}\text{,}\sin\frac{\left(i-9\right)\pi}{2}\Big]\text{,}~9\leq i \leq 12\text{,}\\
		&v_{d}\Big[\cos\frac{\left(2i-9\right)\pi}{4}\text{,}\sin\frac{\left(2i-9\right)\pi}{4}\Big]\text{,}~13\leq i \leq 16\text{,}	
	\end{aligned}
	\right.
\end{equation}
and
\begin{equation}\label{e27}
	\eta_{i}=
	\left\{
	\begin{aligned}
		&\eta_{a}\text{,}~1\leq i\leq4\text{,}\\
		&\eta_{b}\text{,}~5\leq i\leq8\text{,}\\
		&\eta_{c}\text{,}~9\leq i\leq12\text{,}\\
		&\eta_{d}\text{,}~13\leq i\leq16\text{,}
	\end{aligned}
	\right.
\end{equation}
where the discrete parameters $v_{a}$, $v_{b}$, $v_{c}$, $v_{d}$, $\eta_{a}$, $\eta_{b}$, $\eta_{c}$ and $\eta_{d}$ are adjustable.

In order to facilitate calculations, dimensionless variables are employed in this work. The physical quantities are expressed in nondimensional form with reference to the following values, i.e., the molar mass $m_{0}$, molar number density $n_{0}$,  length $L_{0}$, temperature $T_{0}$, and universal gas constant $R$. The details are listed as below
\begin{table}[h]
	\rule{0pt}{15pt}
	\begin{tabular}{l@{\hspace{1em}}l@{\hspace{3em}}l}\vspace{0.3em}
		&Dsitribution funcitions: $f_{i}^{\sigma}$   & by $n_{0}$   \\  \vspace{0.3em}
		&Speed and velocity: $\pmb{v}_{i}$, $\eta_{i}$, $\pmb{u}^{\sigma}$, $\pmb{u}$  &by$\sqrt{RT_{0}/m_{0}}$       \\  \vspace{0.3em}
		&Mass density: $\rho^{\sigma}$, $\rho$        &by $m_{0}n_{0}$       \\  \vspace{0.6em}
		&Energy density: $E^{\sigma}$, $E$   & by $n_{0}RT_{0}$   \\  \vspace{0.3em}
		&Temperature: $T^{\sigma}$, $T$        &by $T_{0}$       \\  \vspace{0.3em}
		&Coordinate: $x$, $y$        &by $L_{0}$       \\  \vspace{0.3em}
		&Distances: $L_{x}$, $L_{y}$        &by $L_{0}$       \\  \vspace{0.3em}
		&Time: $t$   & by $L_{0}/\sqrt{RT_{0}/m_{0}}$   \\  \vspace{0.6em}
	\end{tabular}
\end{table}

\section{Numerical simulations}\label{SecIII}
\begin{figure*}[htbp]
	{\centering
		\includegraphics[width=0.9\textwidth]{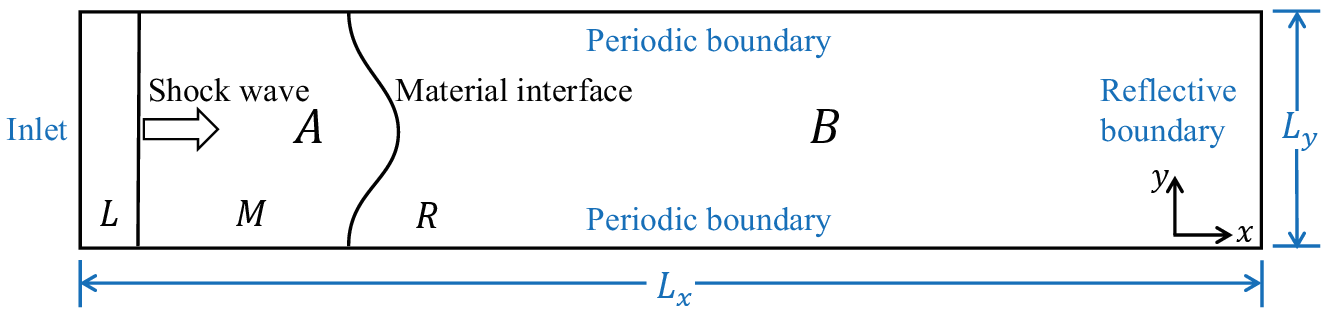}}
	\caption{\centering{Schematic diagram of initial setting of the RM instability.}}
	\label{Fig02}
\end{figure*}

In this section, the RM instability is simulated in a two-dimensional region with space $L_{x}\times L_{y}=0.5\times 0.1$. The initial setup of the numerical simulation is illustrated in Fig. \ref{Fig02}, where the physical quantities in each of these regions are specified as follows
\begin{equation}\label{e28}
	\left\{
	\begin{aligned}
		&\left(\rho, u_{x}, u_{y}, p\right)_{\rm{L}}	= \left(1.3416, 0.3615, 0.0, 1.5133\right)\text{,}\\
		&\left(\rho, u_{x}, u_{y}, p\right)_{\rm{M}} = \left(1.0, 0.0,0.0, 1.0\right)\text{,}\\
		&\left(\rho, u_{x}, u_{y}, p\right)_{\rm{R}} = \left(5.5, 0.0,0.0, 1.0\right)\text{,}	
	\end{aligned}
	\right.
\end{equation}
where the suffix L, M, and R represent three regions in the flow field: the left region ($0\leq x < 0.025$), the middle region ($0.025\leq x < 0.125$), and the right region ($0.125\leq x < 0.5$), respectively. Regions L and M are filled with component $A$, where the particle number density is $n^{A}=1$ and the particle mass is $m^{A}=1$.  In contrast, region R contains component $B$,  with $n^{B}=1$ and $m^{B}=5.5$. Between regions L and M is a shock front traveling rightwards with a Mach number $\rm{Ma}=1.2$. The physical quantities across the shock wave satisfy the Rankine-Hugoniot relations. Meanwhile, between regions M and R is a material interface separating components $A$ and $B$. A sinusoidal perturbation is imposed on the material interface, described by $x=0.125+0.01\cos (2\pi y/L_{y})$.

In addition, the discrete parameters are chosen as $(v_{a}, v_{b}, v_{c}, v_{d}) = (0.6, 1.6, 2.9, 5.9)$ and $(\eta_{a}, \eta_{a}, \eta_{c}, \eta_{d}) = (0.0, 2.9, 0.0, 0.0)$, while the temporal step is set as $\Delta t=5.0\times 10^{-6}$. The left boundary is set as an inflow boundary and the right boundary is treated as a reflective boundary. In the $y$-direction, periodic boundary conditions are applied. To ensure the simulations are both effective and accurate, a grid-independence test is conducted beforehand, with detailed results provided in Appendix \ref{B}. Consequently, a grid resolution of $N_x \times N_y = 2000 \times 400$ is selected for this work, corresponding to a spatial step size of $\Delta x=\Delta y=2.5\times 10^{-4}$.

This work focuses on the impact of relaxation time on the RM instability during the reflection process. The fluid dynamics are analyzed under seven different relaxation parameters, including $\theta ^{A}=\theta ^{B}=2.0\times 10^{-5}$, $4.0\times 10^{-5}$, $8.0\times 10^{-5}$, $1.6\times 10^{-4}$, $3.2\times 10^{-4}$, $6.4\times 10^{-4}$, $1.28\times 10^{-3}$. The formula for Reynolds number is $\textit{Re}=\rho_{c}u_{c}L_{c}/\mu_{c}$. In this case, the average density of components $A$ and $B$ is used as the characteristic density $\rho_{c}=(\rho_{M}+\rho_{R})/2$. The initial shock velocity $u_{x,L}$, representing the $x$-direction velocity in the left region before the interaction of the shock wave with the interface, is selected as the characteristic velocity $u_{c}=u_{x,L}$. The characteristic length is $L_{c}=L_{y}$ and the dynamic viscosity is $\mu_{c}=p_{L}\tau^{\sigma}$. As the relaxation time increases, the Reynolds number decreases. Therefore, the Reynolds numbers corresponding to different relaxation parameters are 11748.75, 5874.38, 2937.19, 1468.59, 734.30, 367.15 and 183.57, respectively.

In the following section, the impact of relaxation time on the RM instability process is explored from three complementary perspectives. First, from the viewpoint of HNE effects, the evolution of RM instability is analyzed, focusing on the changes in the global average density gradient, vorticity (related to velocity gradient) and kinetic energy, using density contours and schlieren images. This analysis reveals the expansion of perturbations at the fluid interface and their influence on the development of instability. Subsequently, the degree of mixing is examined, followed by a geometric analysis of the growth trend of the mixing width over time. This part of the work aims to clarify the role of relaxation time in influencing the mixing behavior of fluids. Finally, the intensity of TNE effects is explored, with an analysis of how relaxation time variations affect TNE intensity within the flow field. This comprehensive analysis enhances our understanding of relaxation time's impact on the RM instability process, providing significant theoretical foundations for future research.

\subsection{Impact of relaxation time on HNE characteristic}
\begin{figure*}[htbp]
	{\centering
		\includegraphics[width=1\textwidth]{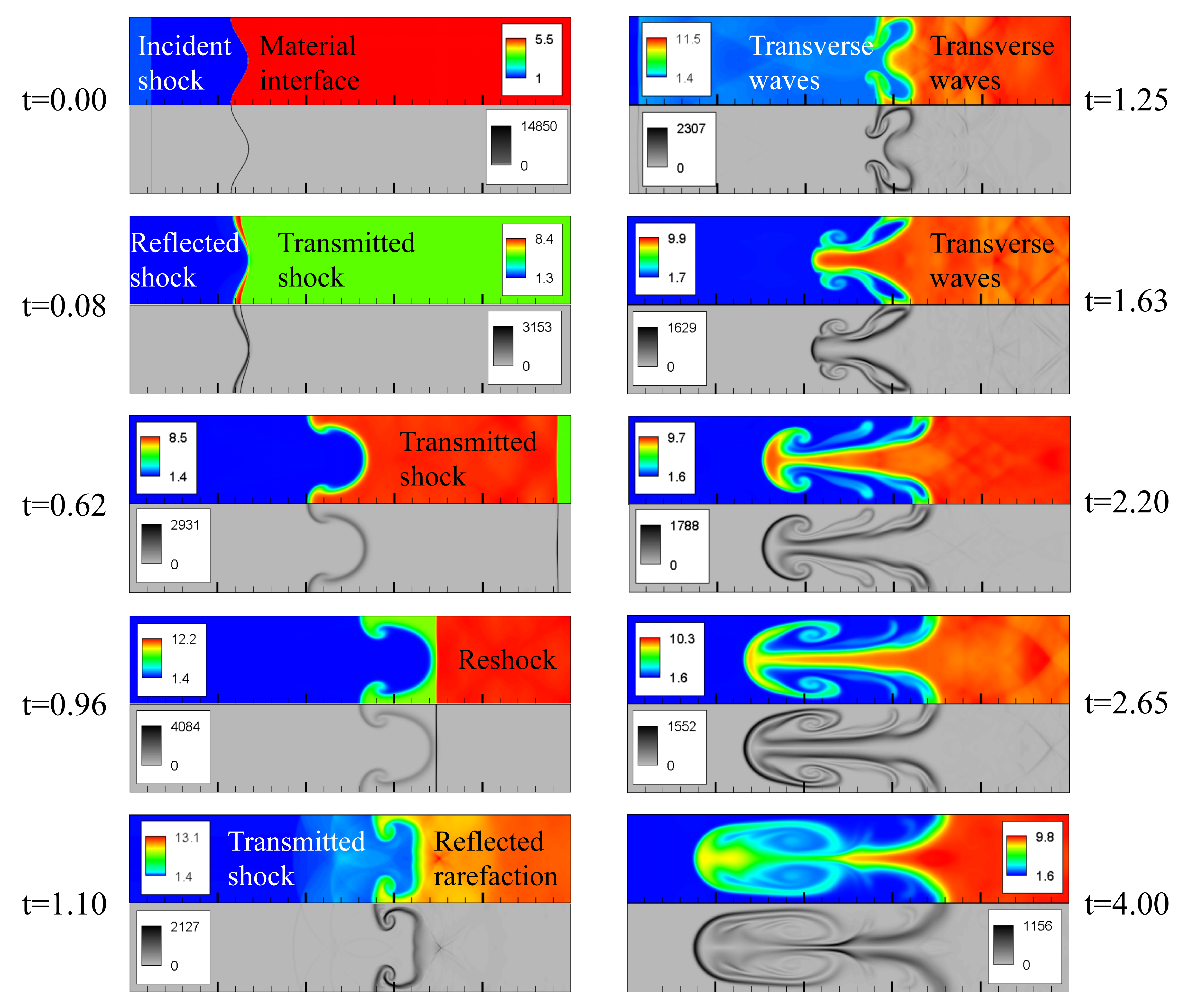}}
	\caption{\centering{Contours of density (top) and schlieren images (bottom) at different moments with a relaxation parameter of $2.0\times 10^{-5}$.}}
	\label{Fig03}
\end{figure*}

To provide a clear understanding of the evolution of the RM instability, Fig. \ref{Fig03} presents the density contours and schlieren images at various moments for a relaxation parameter of $\theta ^{\sigma}=2.0\times 10^{-5}$. The upper portion of each subplot displays the density contours, while the lower portion show the schlieren images of density. At the initial time, the incident shock wave begins propagating rightward and rapidly crosses the material interface at $t=0.08$. This interaction produces a transmitted shock wave propagating through the component $B$, and a reflected shock wave bouncing off the interface. At this stage, the amplitude of the disturbance at the interface grows gradually, although the deformation remains minimal. Subsequently, the two materials on either side of the interface begin to mix. The lighter fluid penetrates the heavier fluid, then the bubble and spike begin to form. Later, the transmitted shock wave reaches the right boundary and reflects back as a reshock wave propagating leftward through the flow field. At $t=0.96$, the reshock wave interacts with the material interface, triggering a secondary RM instability. During this interaction, the reshock wave propagates from the denser fluid into the lighter fluid, suppressing the growth of bubbles. Simultaneously, a reflected rarefaction wave and a transmitted shock wave are generated. The reflected rarefaction wave interacts with the denser fluid, creating a complex wave pattern resembling a ``grid'' structure due to the superposition and interference of disturbances propagating in various directions. Meanwhile, the transmitted shock wave continues moving leftward into the lighter fluid. At $t=1.26$, the transmitted shock wave exits the fluid domain. The reverse acceleration induces further deformation at the material interface, generating of transverse waves on both sides of the interface. This deformation exemplifies the typical nonlinear behavior of RM instability driven by reflected shock waves. With sustained reverse acceleration, the heavy fluid intrudes into the light fluid, causing the peak structures to elongate in the $x$ direction. Over time, velocity shear at the interface generates numerous small-scale vortex structures. These vortices subsequently merge into larger vortices, enhancing the mixing process. Eventually, as the two fluids become fully blended, the vortices dissipate, and the interface becomes indistinct. These simulation results qualitatively agree with previous studies \cite{latini2020comparison}. Furthermore, the schlieren images clearly capture the propagation trajectory of the shock wave, intuitively illustrating the growth direction of interface perturbations and the distribution of velocity shear. These images also highlight the formation of vortex structures during the evolution process. As the system evolves, the schlieren images reveal the gradual decay of these vortices, which eventually disappear at later stages.

\subsubsection{Density gradient}

To better investigate the physical nature underlying the development of the RM instability, the global average density gradient is used in this section to quantify the instability. The mathematical expression for the average density gradient is provided below:
\begin{equation}\label{e31}
	|\overline{\nabla \rho}|=\frac{1}{L_{x}L_{y}}\int_{0}^{L_{x}}\int_{0}^{L_{y}}\sqrt{\Big|\frac{\partial \rho}{\partial x}\Big|^{2}+\Big|\frac{\partial \rho}{\partial y}\Big|^{2}}\mathrm{d}x\mathrm{d}y\text{,}
\end{equation}
where the integral is extended over the whole physical domain. Physically, the average density gradient $|\overline{\nabla \rho}|$ quantifies the spatial non-uniformity of the density field.

\begin{figure}[htbp]
	{\centering
		\includegraphics[width=0.45\textwidth]{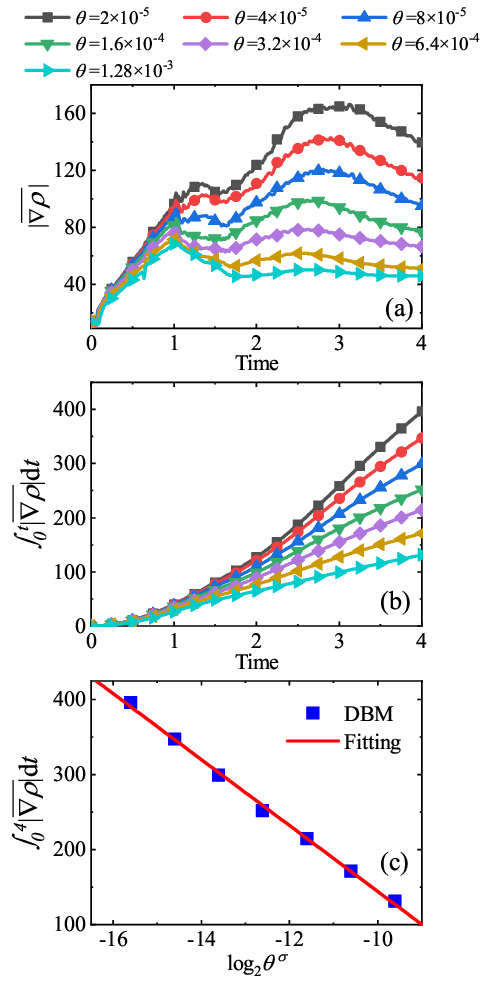}}
	\caption{\centering{(a) Evolution of the average density gradient under different relaxation parameters, (b) growth of the time integral of average density gradient, and (c) the time integrated of the average density gradient at $t = 4.00$ versus the value of $\log_{2}\theta^{\sigma}$.}}
	\label{Fig04}
\end{figure}

Figure \ref{Fig04} (a) demonstrates the variation of the average density gradient over time, with the curves corresponding to different relaxation parameters following a pattern of initially rising, then falling, followed by another rise, and finally either flattening or decreasing. For a more specific analysis, the case where $\theta^{\sigma} =2.0\times 10^{-5}$ is focused on. The shock wave contacts the material interface at $t=0.06$. Then the density gradient is significantly enhanced as there is a reflected shock wave in media $A$ and a transmitted shock wave in media $B$. The growth of the material interface causes a continuous increase in the density gradient until reaches its first peak at $t=1.03$, when the reshock wave contacts the material interface. From $t=1.03$ to $1.55$, the interface is compressed and disturbed by the reshock wave, leading to the interchange of crests and troughs and the formation of transverse waves, the nonlinear characteristics of this process lead to fluctuations in the $|\overline{\nabla \rho}|$. Later, as the heavy fluid intrudes into the light fluid, the interface disturbance intensifies. Simultaneously, the shear effects strengthen the KH instability and promote the generation of vortex structures. This drives the gradual elongation of the spike and bubble structure in the $x$ direction, and the density gradient shows an upward trend. Over time, as the vorticity dissipates and the two components mix, the spike structure gradually disappears, and the density gradient becomes smoother.

Figure \ref{Fig04} (b) displays the time integral growth of the global average density gradient from $t=0.00$ to $4.00$. It can be seen that when the relaxation parameter is small, the $\int_{0}^{t}|\overline{\nabla \rho}|\mathrm{d}t$ increases rapidly, indicating rapid interface evolution and strong disturbance. In contrast, when the relaxation parameter is large, the $\int_{0}^{t}|\overline{\nabla \rho}|\mathrm{d}t$ increases slowly, suggesting that the fluid mixing tends to stabilize. This implies that the larger the relaxation parameter, the more significant the influence of the fluid dissipation mechanism on the interface evolution. Figure \ref{Fig04} (c) shows the relationship between $\int_{0}^{4}|\overline{\nabla \rho}|\mathrm{d}t$ and $\log_{2}\theta^{\sigma}$. The fitting function is $\int_{0}^{4}|\overline{\nabla \rho}|\mathrm{d}t=-44.04\log_{2}\theta^{\sigma}-296.41$. It can be seen that $\int_{0}^{4}|\overline{\nabla \rho}|\mathrm{d}t$ decreases with the increase of $\log_{2}\theta^{\sigma}$. From the physical point of view, for a large relaxation parameter, the diffusion and dissipation become dominant, and the disturbance development of the interface is inhibited, resulting in a small change in physical gradient.

\subsubsection{Vorticity}

To better understand the generation, evolution, and decay of vortex structures in fluids, this section focuses on the global vorticity during the RM instability caused by reshock impact. For a two dimensional system, the mathematical expression for vorticity is as follows
\begin{equation}\label{e30}
	\omega=\frac{\partial u_{y}}{\partial x}-\frac{\partial u_{x}}{\partial y}\text{,}
\end{equation}
which is a function of spatial partial differential of flow velocity.
Then we define the global vorticity as
\begin{equation}
	\langle \omega \rangle=\int_{0}^{L_{x}}\int_{0}^{L_{y}}|\omega|\mathrm{d}x\mathrm{d}y\text{,}
\end{equation}
where the integral is taken over the entire physical domain. The global vorticity $\langle \omega \rangle$ represents the overall intensity of vorticity across the flow field.

\begin{figure}[htbp]
	{\centering
		\includegraphics[width=0.45\textwidth]{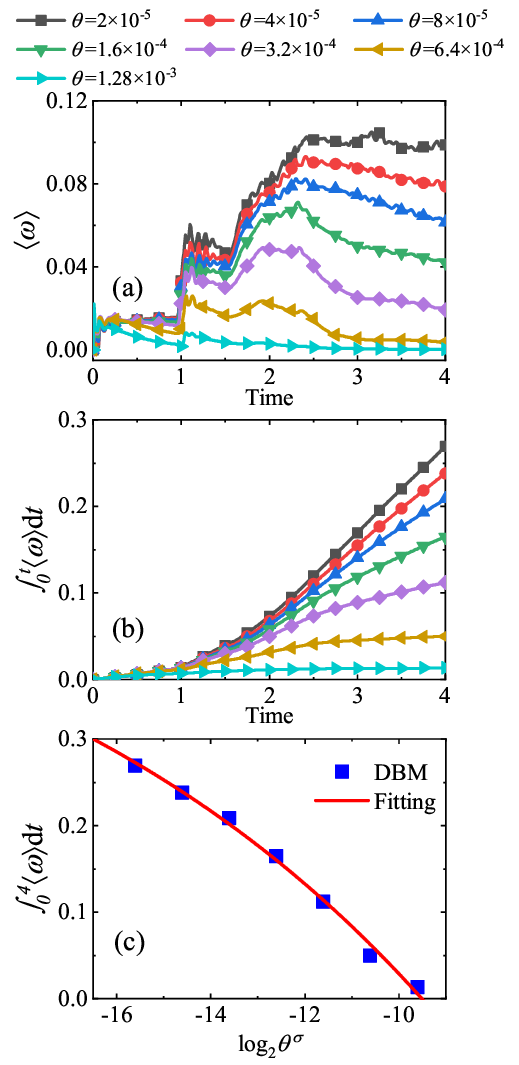}}
	\caption{\centering{(a) Evolution of the global vorticity under different relaxation parameters, (b) growth of the time integral of global vorticity, and (c) the time integrated of the global vorticity at $t = 4.00$ versus the value of $\log_{2}\theta^{\sigma}$.}}
	\label{Fig05}
\end{figure}

Figure \ref{Fig05} (a) illustrates the evolution of the global vorticity with different relaxation parameters. It is evident that for different relaxation parameters, the increasing trend of each curve is roughly the same. Let us take $\theta^{\sigma}=2.0\times 10^{-5}$ as an example. As the shock wave propagates across the material interface, the shear flows are induced around the perturbed interface, which leads to a rapid increase in $\langle \omega \rangle$. During the interval between the first and second impact of the shock wave, the RM instability facilitates the vorticity while the relaxation process suppresses the vorticity, leading the $\langle \omega \rangle$ to enters a relatively smooth development stage. During $t=0.96$ to $1.05$, the reshock wave impacts the material interface for the second time, leading to a sharp increase in $\langle \omega \rangle$ over a short period. Subsequently, the material interface transitions into a highly nonlinear stage, where more complex vortex structures are formed. The interaction between these vortex structures causes the curve of $\langle \omega \rangle$ to oscillate. After $t=1.56$, the fluid velocity difference between the two sides of the material interface increases with time, intensifying the development of KH instability and promoting the generation of vortex structure, which leads to another increase in the $\langle \omega \rangle$. Meanwhile, the continuous stretching of the fluid leads to an increase in the width of the mixed layer, which, in turn, results in the sustained increase in $\langle \omega \rangle$. Finally, as dissipation dominates, the mixed layer tends to saturate, causing the growth of $\langle \omega \rangle$ to gradually slow down and eventually stabilize or decline.

In addition, by comparing the $\langle \omega \rangle$ under different relaxation parameters, it can be found that during the period when the shock wave propagates through component $B$ (approximately from time $0.08$ to $0.96$), the curve of $\langle \omega \rangle$ becomes progressively smoother and even exhibits a noticeable decline as the relaxation parameter increases. This can be attributed to the fact that, with a larger relaxation parameter, the fluid’s viscosity increases, which in turn suppresses the RM instability. As a result, the energy required for vortex formation become insufficient, making it more difficult for the vortex structures to persist, and in some cases, they even weaken. Following the reshock, $\langle \omega \rangle$ exhibits a more rapid decay, further reflecting the influence of the relaxation parameter on the instability evolution.

Figure \ref{Fig05} (b) shows the time integral of the global vorticity from $t=0.00$ to $4.00$, with all curves showing an upward trend. As the relaxation parameter increases, the $\langle \omega \rangle$ grows more slowly, indicating that a larger relaxation parameter more strongly inhibits the RM instability and the generation of vortex structures. Figure \ref{Fig05} (c) depicts the relationship between $\int_{0}^{4}\langle \omega \rangle \mathrm{d}t$ and $\log_{2}\theta^{\sigma}$. The fitting function is given by $\int_{0}^{4}\langle \omega \rangle \mathrm{d}t=-1.57\exp(0.11\log_{2}\theta^{\sigma})+0.58$. It is observed that $\int_{0}^{4}\langle \omega \rangle \mathrm{d}t$ decreases exponentially as $\log_{2}\theta^{\sigma}$ increases. This shows that the larger the relaxation parameter, the slower the evolution of the RM instability.

\subsubsection{Kinetic energy}
Kinetic energy is a fundamental physical quantity that plays a crucial role in characterizing fluid dynamics, providing essential insights into the process of instability evolution. Mathematically, the formula for the total kinetic energy is expressed by
\begin{equation}
	\langle E_{k} \rangle=\int_{0}^{L_{x}}\int_{0}^{L_{y}}\frac{1}{2}\rho u^{2}\mathrm{d}x\mathrm{d}y\text{,}
\end{equation}
where the integral is performed over the entire physical domain. The total kinetic energy is used to describe the total kinetic energy over the entire computational domain.

\begin{figure*}[htbp]
	{\centering
		\includegraphics[width=0.9\textwidth]{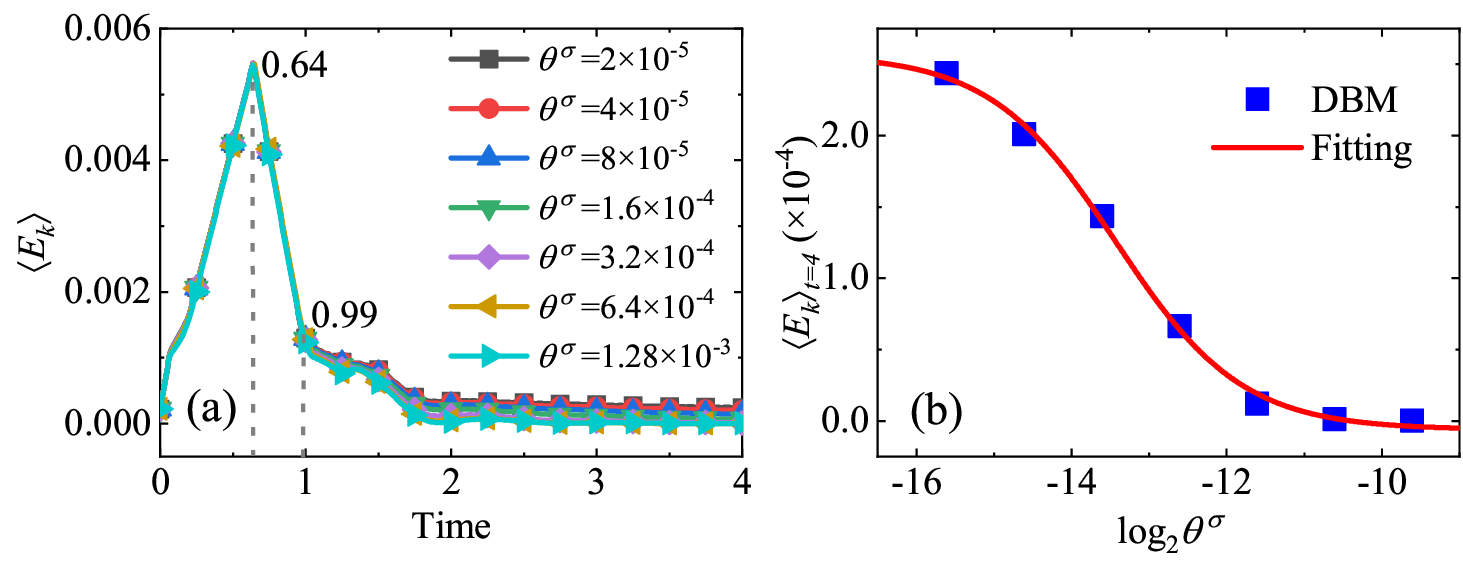}}
	\caption{\centering{(a) Evolution of the total kinetic energy under different relaxation parameters, (b) the relationship between $\log_{2}\theta^{\sigma}$ and $\langle E_{k} \rangle$ at $t=4.00$.}}
	\label{Fig06}
\end{figure*}

Figure \ref{Fig06} (a) illustrates the evolution of kinetic energy under different relaxation parameters. The curve exhibits distinct stage characteristics, initially rising and then declining. During the rising period (from $t=0.00$ to $t=0.64$), the shock wave firstly contacts the material interface (around $t=0.06$), then a reflected shock wave travels leftwards, and a transmitted shock wave propagates rightwards (until $t=0.64$). Because of the right-moving shock wave, the fluid flows rightwards with high kinetic energy, resulting in a gradual increase in total kinetic energy over time. During the declining period ($t>0.64$), after the shock wave reaches the right boundary of the computational domain at $t=0.64$, the reshock wave interacts with the left-flowing fluid, decelerating the flow velocity to near zero, resulting in a reduction of the overall kinetic energy. Notably, prior to the second interaction with the material interface at $t=0.99$, the total kinetic energy curves under different relaxation parameters exhibit remarkable consistency. However, after $t=0.99$, as the interface undergoes the secondary impact, the curve begins to diverge. Specifically, larger relaxation parameters correspond to lower total kinetic energy. This behavior can be attributed to the increased dissipation of kinetic energy into internal energy as the relaxation parameter grows.

Figure \ref{Fig06} (b) shows the relationship between $\langle E_{k} \rangle$ at $t=4.00$ and $\log_{2}\theta^{\sigma}$. The fitting function is given by: $\langle E_{k} \rangle(t=4.00)=\{-6.15+2.51/\left[1+\exp \left(1.23\log_{2}\theta^{\sigma}+16.58\right)\right]\}\times 10^{-4}$. Obviously, the kinetic energy decreases as the relaxation parameter increases. Specifically, the kinetic energy approaches zero at a higher value of the relaxation parameter. This behavior can be attributed to the interplay between the relaxation parameter and the system's dynamics. As the relaxation parameter increases, the formation of vortices and the evolution of RM instability are slowed, resulting in greater dissipation of kinetic energy into internal energy and, consequently, a reduction in total kinetic energy.

\subsection{Impact of relaxation time on mixing layer}

The degree of mixing is next quantitatively analyzed, and the temporal evolution of the mixing width is examined from a geometric perspective. As the RM instability progresses, the mixing layer gradually expands. The change in the mixing width reflects the degree of mixing between the two fluids. By precisely tracking this evolution, the influence of relaxation parameter on the mixing efficiency between the fluids can be revealed.

\subsubsection{Mixing degree}

In this work, the mixing degree is defined as
\begin{equation}\label{e29}
	M=4 X^{A}X^{B}\in \left[0\text{,}1\right]\text{.}
\end{equation}
Here, $X^{\sigma}=n^{\sigma}/(n^{A}+n^{B})$ represents the concentrations of components $A$ or $B$. When $X^{A}$ and $X^{B}$ are completely separated (i.e.,  $X^{A}=0$ or  $X^{B}=0$), the mixing degree $M=0$. When  $X^{A}$ and  $X^{B}$ are fully mixed uniformly (i.e.,  $X^{A}= X^{B}=0.5$), the mixing degree reaches its maximum value of $M=1$.

\begin{figure*}[htbp]
	{\centering
		\includegraphics[width=1\textwidth]{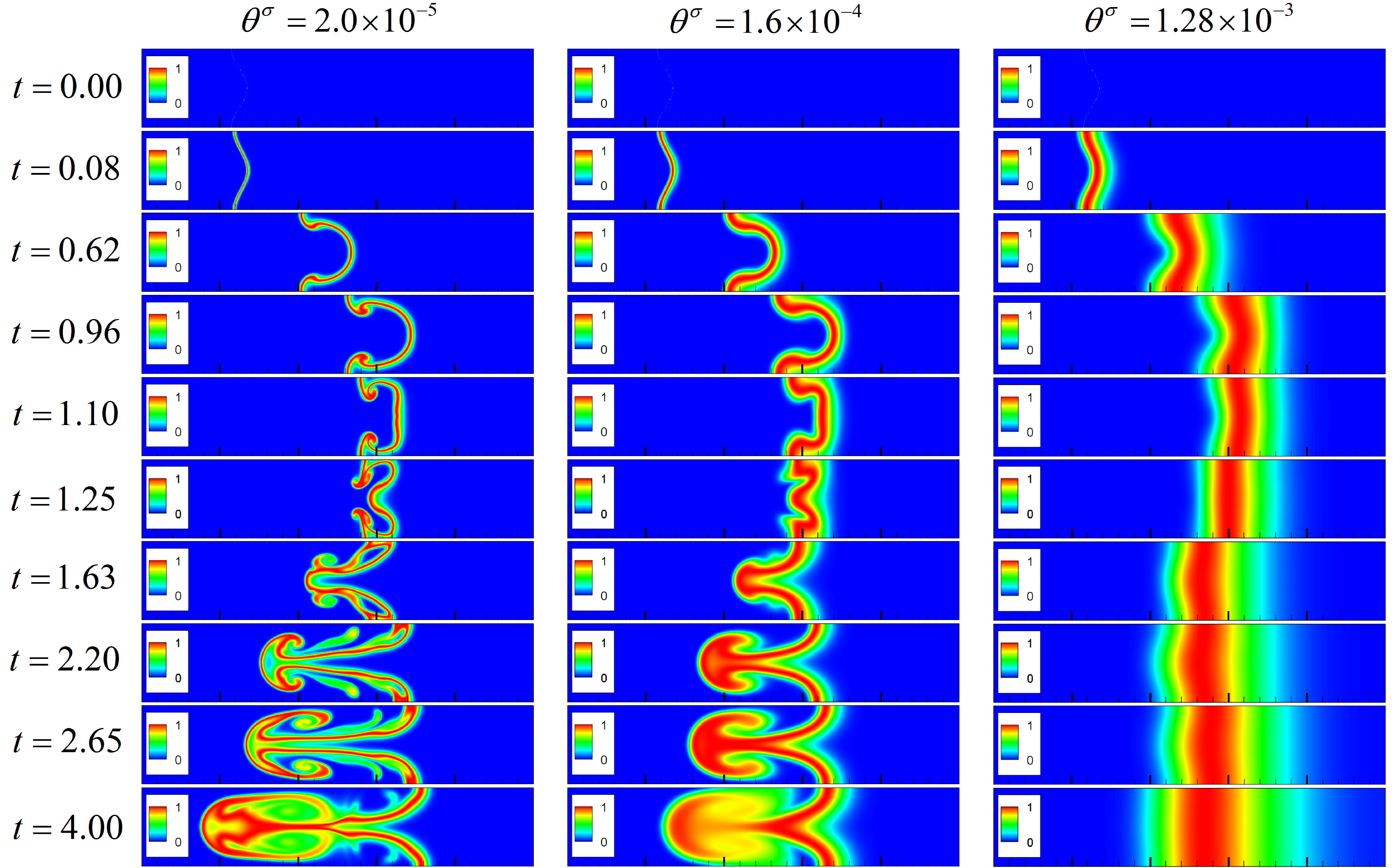}}
	\caption{\centering{Evolution of mixing degree for three relaxation parameters: $\theta^{\sigma}$ = $2.0\times 10^{-5}$, $1.6\times 10^{-4}$, and $1.28\times 10^{-3}$.}}
	\label{Fig07}
\end{figure*}

To intuitively assess the mixing state of the fluids, Fig. \ref{Fig07} shows the mixing degree contours in the RM process. From left to right, three columns are for $\theta^{\sigma} = 2.0\times 10^{-5}$, $1.6\times 10^{-4}$ and $1.28\times 10^{-3}$, respectively. Let us focus on the first column as an example. Initially, there is no mass diffusion on either side of the material interface, and the fluid is in a non-mixed state. After the incident shock wave impacts the material interface, the perturbed interface grows, thereby enlarging the region where intense mixing occurs. However, the overall degree of mixing remains low at this stage. After the reshock wave passes through the material interface again, the region of mixed layer expands significantly. The density and velocity gradients triggers the KH instability near the interface. This instability manifests as complex vortex structures that further amplify the fluid instability. At the same time, the heavy fluid gradually intrudes into the light fluid, and the width of the mixed layer increases rapidly due to the development of RM instability, which is manifested as a significant expansion of the mixed regions.

Furthermore, a comparative analysis of mixing degree contours is conducted across various relaxation parameters. It is evident that structural features such as bubbles and spikes are clearly observed when the relaxation parameter is small. As the relaxation parameter increases, the mixing layer becomes wider without fine structures around the material interface. The physical reasons are as follows. On the one hand, the increase in the relaxation parameter enhances the diffusion effect, causing the interface thickness to grow rapidly while gradually reducing the density difference across the interface. This lowers the Atwood number, thereby suppresses further development of the RM instability. On the other hand, a larger relaxation time results in a higher dynamic viscosity coefficient, which reduces the velocity gradient near the interface under the influence of shear forces. This further inhibits the onset of KH instability, preventing the formation of vortex structures in the later stages. Ultimately, fine-scale features are suppressed, and the interface becomes smooth. These phenomena underscore the significant role of the relaxation parameter in suppressing RM instability.

\subsubsection{Mixing width}

The evolution of mixing width during the development of RM instability is subsequently analyzed, focusing on the continuous interaction and gradual merging processes between light and heavy fluids, which lead to the formation of a distinct mixing region. To quantify the extent of this mixing, we define the interface width ($W_I$) as the horizontal distance between the leftmost and rightmost positions where the mixing degree reaches $0.9$, and the mixed layer width ($W_M$) as the distance between the positions where the mixing degree is $0.1$.

\begin{figure*}[htbp]
	{\centering
		\includegraphics[width=0.85\textwidth]{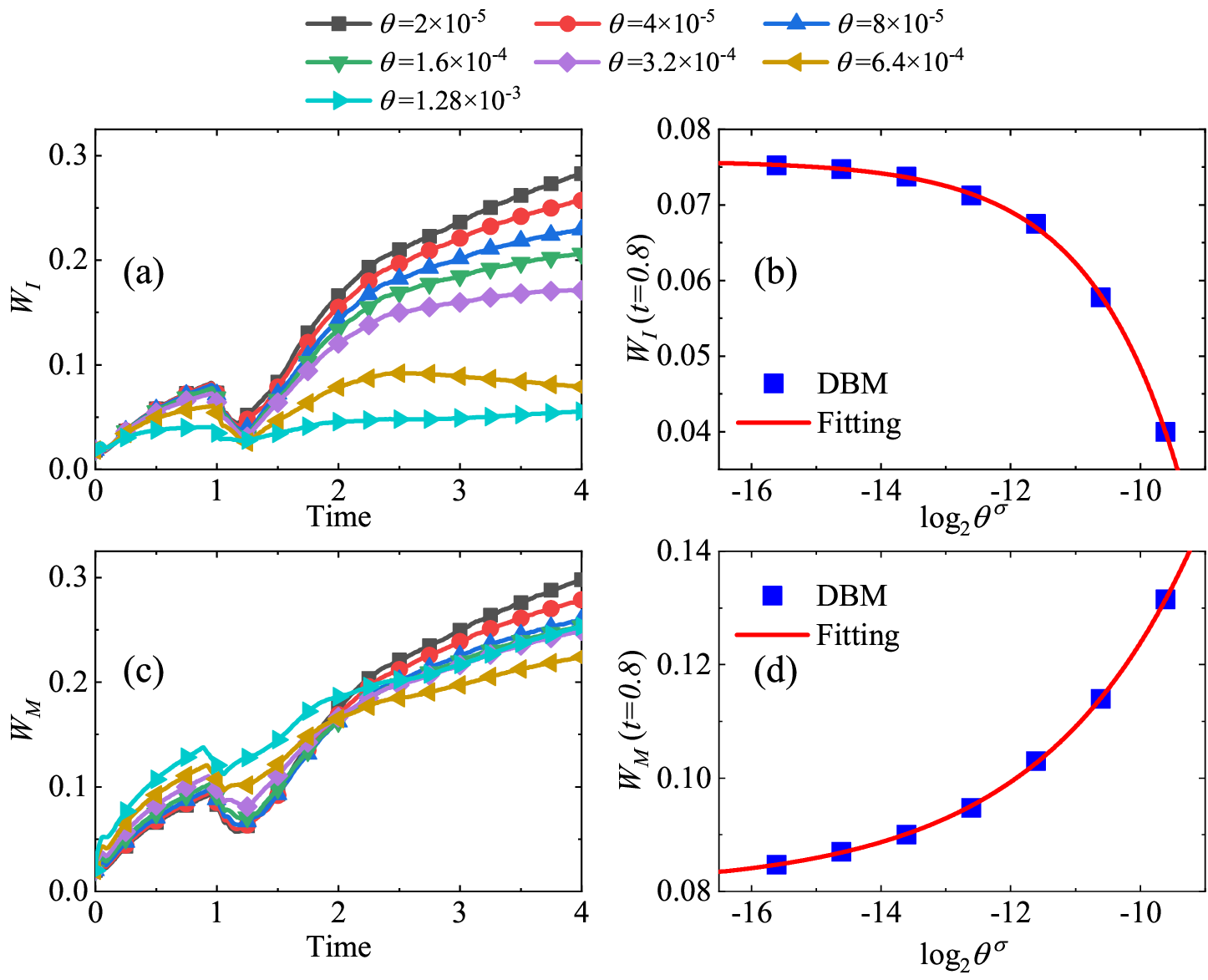}}
	\caption{\centering{Evolution of the interface width (a) and mixed layer width (c) under different relaxation parameters, the relationship between $\log_{2}\theta^{\sigma}$ and the interface width (b) and mixed layer width (d) at $t = 0.8$.}}
	\label{Fig08}
\end{figure*}

Figure \ref{Fig08} (a) illustrates the evolution of the interface width $W_I$ over time for different relaxation parameters. All curves exhibit a pattern of initial increase, followed by a decrease, and a final increase. Take $\theta^{\sigma} = 2 \times 10^{-5}$ as an example, from $t = 0.00$ to $0.08$, there is an oscillation due to the diffusion of the components and the compression of the shock wave on the material interface. Afterward, when the transmitted shock wave passes through component $B$, the disturbance at the interface is amplified by the RM instability, causing the interface width to increase again. At $t = 0.96$, the reshock wave interacts with the material interface for the second time, and the reverse acceleration inhibits interface growth, resulting in a decrease in the interface width. By $t = 1.17$, the peak and valley shift into each other, and the interface width reaches its minimum. Subsequently, as the RM instability develops, the interface width enters a phase of rapid expansion. However, as the two fluids progressively approach a fully mixed state, the growth rate of the interface width diminishes. Overall, a larger relaxation parameter leads to a smaller interface width, suggesting that higher relaxation parameters inhibit the development of RM instability.

Figure \ref{Fig08} (c) illustrates the temporal evolution of the mixed layer width $W_M$ under varying relaxation parameters. It is noteworthy that, in the early stages, an increase in the relaxation parameter enhances diffusion, thereby accelerating the mixing process and leading to an increase in the mixed layer width. However, in the final stage, the mixed layer width shows a complex change with the relaxation parameter. For example, at the time $t=4.00$, it first decreases and then increases as the relaxation parameter increases. Specifically, at smaller relaxation parameters, the mixed layer width gradually decreases with increasing relaxation parameter. Yet, as the relaxation parameter continues to rise, the mixed layer width begins to expand, demonstrating a clear reversal in trend. This behavior occurs because an increase in the relaxation parameter not only reduces the density gradient near the interface, thereby mitigating RM instability, but also diminishes the velocity gradient, which subsequently inhibits KH instability. Consequently, the mixed layer width initially decreases. However, as the relaxation effect becomes dominant with further increases in the relaxation parameter, the diffusion of particle intensifies, leading to a corresponding increase in the mixed layer width.

Figures \ref{Fig08} (b) and (d) present the relationship between $\log_{2}\theta^{\sigma}$ and the interface width and mixed layer width at $t = 0.8$. The fitting function are $W_{I}(t=0.8)=-30.97\exp \left(0.70\log_{2}\theta^{\sigma}\right)+0.08$ and $W_{M}(t=0.8)=2.86\exp \left(0.42\log_{2}\theta^{\sigma}\right)+0.08$, respectively. These observations reflect the evolution of both the interface and mixed layer widths under the initial shock wave, with the two exhibiting opposite trends as the relaxation parameter increases. This behavior can be attributed to the distinct effects of the relaxation parameter on the RM instability. As the relaxation parameter increases, the density gradient near the interface diminishes, thereby suppressing RM instability and leading to a contraction of the interface width. At the same time, the increase in the relaxation parameter intensifies diffusion, which promotes the growth of the mixed layer, causing its width to gradually expand.

\subsection{Impact of relaxation time on TNE characteristic}

\begin{figure*}[htbp]
	{\centering
		\includegraphics[width=0.85\textwidth]{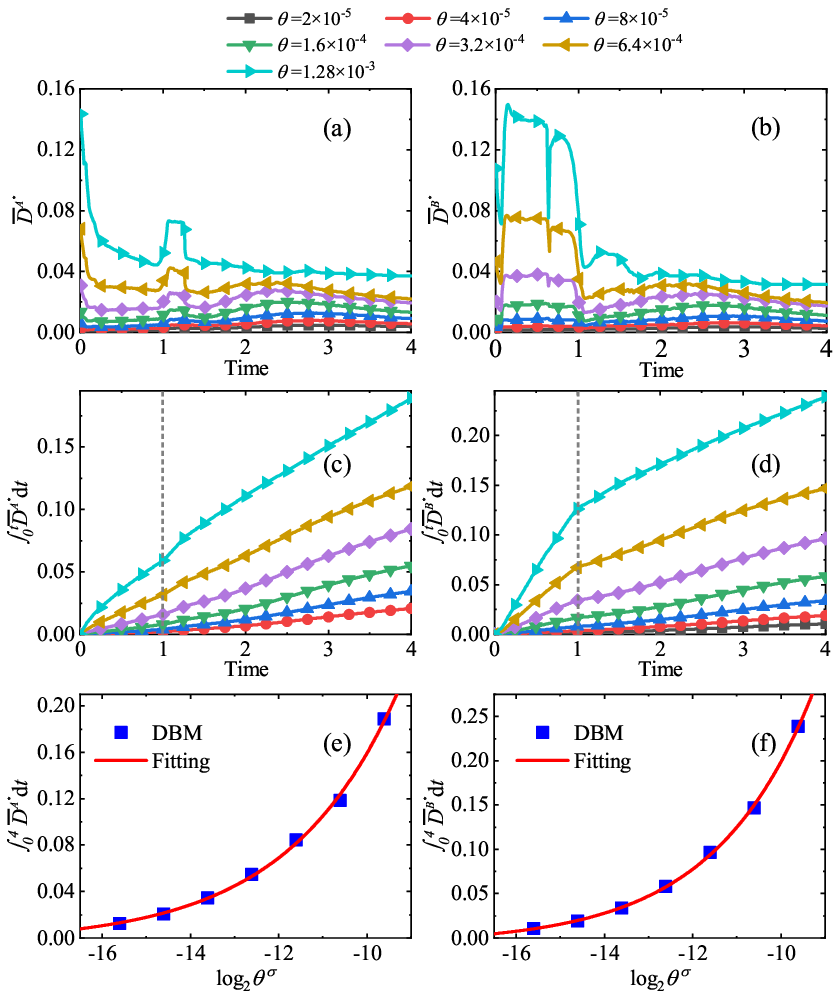}}
	\caption{\centering{Global average TNE intensity for components $A$ (a) and $B$ (b) at different relaxation parameters; time integral growth of the global average TNE intensity for components $A$ (c) and $B$ (d); and time integral of the global average TNE intensity at $t=4.00$ for components $A$ (e) and $B$ (f) as a function of $\log_{2}\theta^{\sigma}$.}}
	\label{Fig09}
\end{figure*}

We now investigate the influence of varying relaxation times on RM instability from the perspective of TNE.
Figures \ref{Fig09} (a) and (b) illustrate the evolution of the average TNE intensity for components $A$ and $B$, respectively. Both $\overline{D}^{A^{*}}$ and $\overline{D}^{B^{*}}$ demonstrate a gradual increase with increasing relaxation parameter. This observed trend can be physically explained by the fact that systems with smaller relaxation parameters exhibit faster recovery from non-equilibrium states to thermodynamic equilibrium.

The evolution of TNE manifestations is governed by three competing physical mechanisms: (i) Diffusion and dissipation, which smooth the gradients of density and velocity, leading to a decrease in $\overline{D}^{\sigma^{*}}$. (ii) RM and KH instabilities, which enhance interface stretching and increase physical quantity gradients, thereby driving an increase in $\overline{D}^{\sigma^{*}}$. (iii) Shock waves, rarefaction waves, and transverse waves, which maintain physical gradients and consequently enhance $\overline{D}^{\sigma^{*}}$. The interplay of these three mechanisms results in the complex TNE characteristics observed in Figs. \ref{Fig09} (a) and (b), respectively.

As shown in Fig. \ref{Fig09} (a), $\overline{D}^{A^{*}}$ exhibits distinct evolutionary phases. From $t=0.00$ to about $t=0.25$, $\overline{D}^{A^{*}}$ decreases due to the dominance of the first mechanism. Later, from $t=0.25$ to around $t=0.96$, $\overline{D}^{A^{*}}$ increases for smaller relaxation parameters as the second mechanism becomes dominant, whereas it continues to decrease for larger relaxation parameters due to the prevailing influence of the first mechanism. From $t=0.96$ to $t=1.07$, $\overline{D}^{A^{*}}$ shows a significant increase due to the fact that the reshock wave passes through the material interface and enters component $A$, enhancing the mixing of the components around the interface. Between $t=1.07$ and $t=1.28$, $\overline{D}^{A^{*}}$ fluctuates due to the competition of the first and second mechanisms. Then it falls quickly once the transmitted shock wave and transverse waves exit the fluid domain. When $t>1.52$, $\overline{D}^{A^{*}}$ shows different evolutionary trends depending on the relaxation parameter. To be specific, for smaller relaxation parameters, the curve initially increases and then decreases because of the competing first and second mechanisms. However, for a specifically large relaxation parameter, such as $\theta^{\sigma}=1.28\times 10^{-3}$, $\overline{D}^{A^{*}}$ exhibits a consistent decline, because the diffusion dominates the process.

In Fig. \ref{Fig09} (b), $\overline{D}^{B^{*}}$ displays the following evolutionary characteristics. At $t=0.08$, $\overline{D}^{B^{*}}$ undergoes a rapid surge in all cases due to the interaction between the shock wave and the material interface, which significantly enhances the physical gradient.
From $t=0.08$ to $t=0.64$, $\overline{D}^{B^{*}}$ increases for smaller relaxation parameters as the third mechanism dominates, whereas it decreases for larger relaxation parameters due to the dominance of the first mechanism. Between $t=0.64$ and $t=1.07$, the shock wave is reflected and propagates leftward, passing through the material interface and leaving component $B$. This behavior is primarily related to the third mechanism. For $t>1.07$, $\overline{D}^{B^{*}}$ exhibits an oscillatory pattern characterized by an initial increase followed by a decrease, resulting from the competing influences of the first and second mechanisms.

Figures \ref{Fig09} (c) and (d) show the growth of the time integral of the global average TNE intensity for components $A$ and $B$, respectively. Clearly, both exhibit an upward trend and increase with the relaxation parameter. Notably, around $t=1.00$, the secondary shock induces significant variations in the TNE intensity of components $A$ and $B$, resulting in a distinct inflection point in the time integral curve of TNE intensity.
Furthermore, Figs. \ref{Fig09} (e) and (f) illustrate the relationship between $\log_{2}\theta^{\sigma}$ and $\int_{0}^{4}\overline{D}^{\sigma^{*}}\mathrm{d}t$ , where the fitting functions are given by $\int_{0}^{4}\overline{D}^{A^{*}}\mathrm{d}t=9.11\exp \left(0.40\log_{2}\theta^{\sigma}\right)-0.01$ and $\int_{0}^{4}\overline{D}^{B^{*}}\mathrm{d}t=17.75\exp \left(0.45\log_{2}\theta^{\sigma}\right)-0.01$, respectively. These results indicate that the TNE intensity exhibits a monotonic increase with increasing relaxation parameters.

\section{Conclusions}\label{SecIV}

In this study, a two-component DBM is utilized to simulate the evolution of RM instability under varying relaxation times. The influence of relaxation time on HNE characteristics, mixed layer evolution, and TNE behavior is systematically investigated.

First, the HNE behavior during RM instability is analyzed. The results demonstrate that increasing relaxation time enhances diffusive and dissipative mechanisms, thereby suppressing RM instability and leading to a gradual reduction in the global average density gradient. Furthermore, larger relaxation times inhibit the formation of vortex structures, resulting in a significant decrease in vorticity. The temporal evolution of kinetic energy further supports this mechanism, revealing a decrease in system kinetic energy with increasing relaxation time. This observation confirms the role of relaxation time in suppressing instability and decelerating the development of fluid disturbances.

Second, the effects of relaxation time on mixing degree, interfacial width, and mixed layer width are examined in detail. As relaxation time increases, characteristic fluid structures (e.g., spikes, bubbles, and vortices) gradually diminish. Notably, the interface width and mixed layer width exhibit opposing trends during the early stages of evolution. The interface width decreases due to the suppression of RM instability, while the mixed layer width increases as enhanced diffusion promotes greater inter-component mixing. In later stages, further increases in relaxation time suppress both RM and KH instabilities, causing the mixed layer width to decrease. However, when diffusion becomes the dominant mechanism, the mixed layer width increases again due to the continued enhancement of diffusive effects.

Finally, the TNE behavior is investigated. The results indicate that the global averaged non-equilibrium intensity increases with relaxation time but exhibits distinct evolutionary trends depending on the relaxation time. This behavior arises from the competition among multiple physical mechanisms, including diffusion, dissipation, RM instability, KH instability, shock waves, rarefaction waves, and transverse waves. These mechanisms interact in a coupled manner at different stages of the flow evolution, leading to variable TNE manifestations as relaxation time changes. These findings provide significant insights into the underlying physical mechanisms of compressible fluid systems.

\begin{acknowledgments}
This work is supported by National Natural Science Foundation of China (under Grant No. U2242214), Guangdong Basic and Applied Basic Research Foundation (under Grant No. 2024A1515010927), Humanities and Social Science Foundation of the Ministry of Education in China (under Grant No. 24YJCZH163), Fujian Provincial Units Special Funds for Education and Research (2022639), and Fundamental Research Funds for the Central Universities, Sun Yat-sen University (under Grant No. 24qnpy044). This work is partly supported by the Open Research Fund of Key Laboratory of Analytical Mathematics and Applications (Fujian Normal University), Ministry of Education, P. R. China (under Grant No. JAM2405).
\end{acknowledgments}

\appendix
\section{Hydrodynamic equations}\label{A}
Using the Chapman--Enskog expansion, Eq. (\ref{DBM}) can be reduced to the NS equations in the continuum limit \cite{lin2016double}, as follows:
\begin{equation}
	\label{NSsigma-1}
	\frac{\partial {{\rho }^{\sigma }}}{\partial t}+\frac{\partial }{\partial {{r}_{\alpha }}}\left( {{\rho }^{\sigma }}u_{\alpha }^{\sigma } \right)=0
	\text{,}
\end{equation}%
\begin{eqnarray}
	\label{NSsigma-2}
	& \dfrac{\partial }{\partial t}\left( {{\rho }^{\sigma }}u_{\alpha }^{\sigma } \right)+\dfrac{\partial }{\partial {{r}_{\beta }}}\left( {{\delta }_{\alpha \beta }}{{p}^{\sigma }}+{{\rho }^{\sigma }}u_{\alpha }^{\sigma }u_{\beta }^{\sigma }\right)+\dfrac{\partial }{\partial {{r}_{\beta }}}\left(P_{\alpha \beta }^{\sigma }+U_{\alpha \beta }^{\sigma }\right)\nonumber\\ [8pt]
	&=-\dfrac{\rho^{\sigma }}{\tau^{\sigma}}\left(u_{\alpha}^{\sigma}-u_{\alpha}\right)\text{,}
\end{eqnarray}%
\begin{eqnarray}
	\label{NSsigma-3}
	& \dfrac{\partial }{\partial t}\left[{{\rho }^{\sigma }}\left( {{e}^{\sigma }}+\dfrac{1}{2}{{u}^{\sigma 2}} \right)\right]+\dfrac{\partial }{\partial {{r}_{\alpha }}}\left[ {{\rho }^{\sigma }}u_{\alpha }^{\sigma }\left( {{e}^{\sigma }}+\dfrac{1}{2}{{u}^{\sigma 2}} \right)+{{p}^{\sigma }}u_{\alpha }^{\sigma } \right] \nonumber \\ [8pt]
	& -\dfrac{\partial }{\partial {{r}_{\alpha }}}\left[ {{\kappa }^{\sigma }}\dfrac{\partial {T}^{\sigma }}{\partial {{r}_{\alpha }}}-u_{\beta }^{\sigma }P_{\alpha \beta }^{\sigma }+Y_{\alpha }^{\sigma }\right]\nonumber \\ [8pt]
	&=-\dfrac{\rho^{\sigma }}{\tau^{\sigma}}\left(\dfrac{D+I}{2} \dfrac{T^{\sigma}}{m^{\sigma}}+\dfrac{1}{2}u^{\sigma 2}-\dfrac{D+I}{2} \dfrac{T}{m^{\sigma}}-\dfrac{1}{2}u^{\sigma} \right)\text{,}
\end{eqnarray}
with
\begin{equation}
	P_{\alpha \beta }^{\sigma }=-{{\mu }^{\sigma }}\left( \frac{\partial u_{\alpha }^{\sigma }}{\partial {{r}_{\beta }}}+\frac{\partial u_{\beta }^{\sigma }}{\partial {{r}_{\alpha }}}-\frac{2{{\delta }_{\alpha \beta }}}{D+I}\frac{\partial u_{\chi }^{\sigma }}{\partial {{r}_{\chi }}} \right)\text{,}
\end{equation}
\begin{equation}
	U_{\alpha \beta }^{\sigma }=-\rho^{\sigma}\left[\dfrac{\delta_{\alpha \beta}}{D+I}\left( u^{\sigma 2}+u^{2}-2u_{\chi}^{\sigma}u_{\chi} \right) \right]\text{,}
\end{equation}
\begin{eqnarray}
	&Y_{\alpha }^{\sigma }=\dfrac{\rho^{\sigma}u_{\alpha}^{\sigma}}{D+I}\left( u_{\beta}^{\sigma}-u_{\beta} \right)^{2}+\rho^{\sigma}\left(u_{\alpha}^{\sigma}-u_{\alpha} \right)\nonumber \\ [8pt]
	 &\times\left( -\dfrac{D+I+2}{2}\dfrac{T^{\sigma}-T}{m^{\sigma}}-\dfrac{1}{2}u^{\sigma 2}+\dfrac{1}{2}u^{2}\right)\text{.}
\end{eqnarray}
Here, the pressure $p^{\sigma }=n^{\sigma}T^{\sigma }$, the internal energy per unit mass $e^{\sigma }=(D+I)T^{\sigma }/(2m^{\sigma })$, the dynamic viscosity coefficient $\mu ^{\sigma }=p^{\sigma }\tau ^{\sigma }$, the heat conductivity $\kappa^{\sigma}=\gamma\mu ^{\sigma }$ and the specific heat ratio $\gamma=(D+I+2)/(D+I)$.

\section{Grid independence}\label{B}
\begin{figure}[htbp]
	{\centering
		\includegraphics[width=0.4\textwidth]{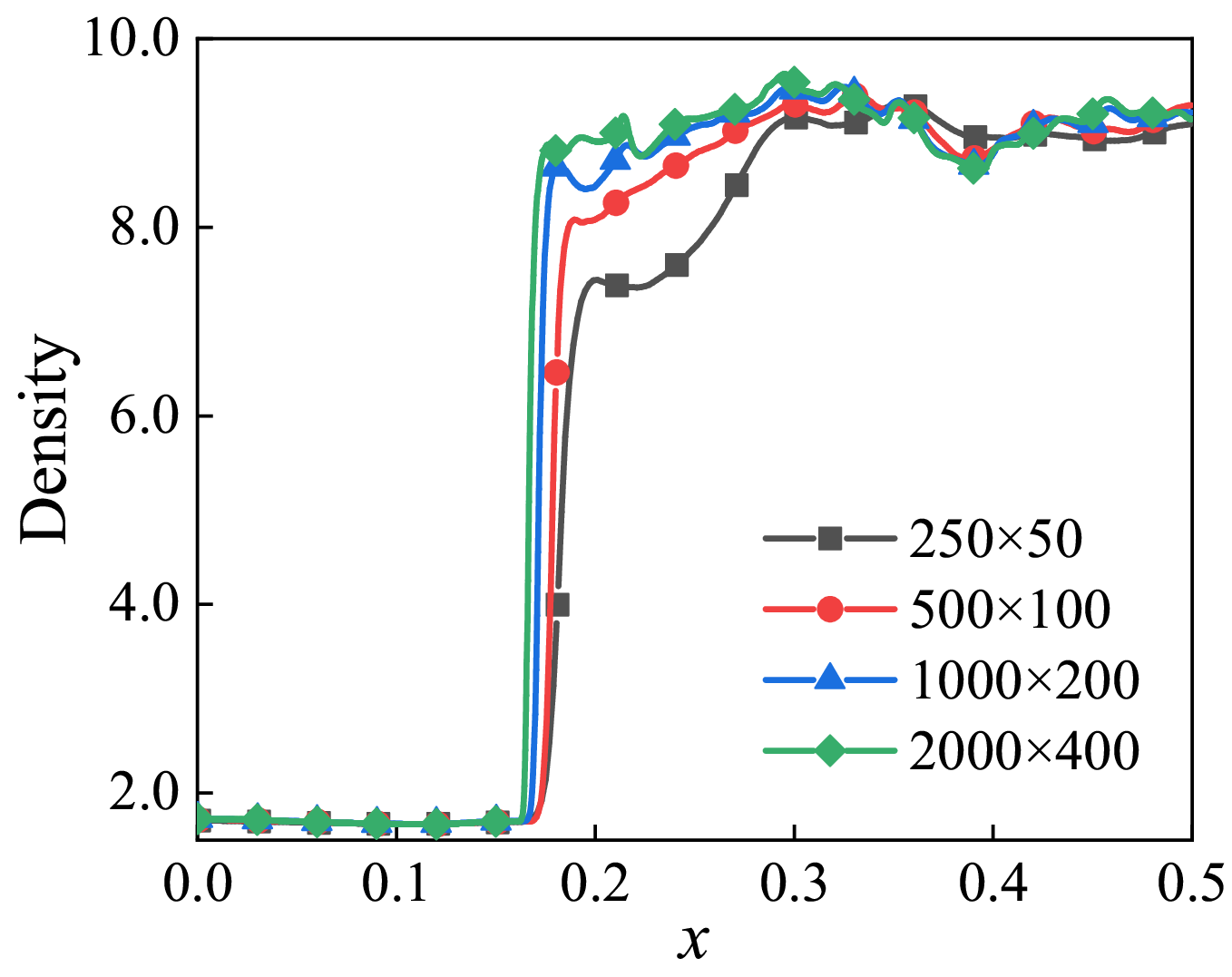}}
	\caption{\centering{Grid independence test: Density profiles along the centerline at a specific time instant during the RM process. Numerical results are plotted for four different mesh resolutions.}}
	\label{Fig10}
\end{figure}

To ensure the accurate simulation of the RM instability, a grid independence test is conducted. Four different grid resolutions are examined with $N_{x}\times N_{y}=250\times 50, 500\times 100, 1000\times 200$, and $2000\times 400$, the corresponding space steps are $\Delta x=\Delta y=2.0\times 10^{-3}, 1.0\times 10^{-3}, 5.0\times 10^{-4}$ and $2.5\times 10^{-4}$, respectively.
Figure \ref{Fig10} presents the density distribution along the centerline $L_{y}/2$ at $t=2.0$ under the condition $\theta ^{A}=\theta ^{B}=2.0\times 10^{-5}$. It can be found that the simulation results progressively converge as the grid resolution increases. The simulated results of $1000\times 200$ and $2000\times 400$ are close to each other. Therefore, a grid resolution of $2000\times 400$ with a space step of $\Delta x=\Delta y=2.5\times 10^{-4}$ is adopted in this work as a reliable computational setup.



\section*{Data Availability}
The data that support the findings of this study are available from the corresponding author upon reasonable request.

\section*{References}
\bibliography{DDBM_RMI}

\begin{thebibliography}{75}%
\makeatletter
\providecommand \@ifxundefined [1]{%
 \@ifx{#1\undefined}
}%
\providecommand \@ifnum [1]{%
 \ifnum #1\expandafter \@firstoftwo
 \else \expandafter \@secondoftwo
 \fi
}%
\providecommand \@ifx [1]{%
 \ifx #1\expandafter \@firstoftwo
 \else \expandafter \@secondoftwo
 \fi
}%
\providecommand \natexlab [1]{#1}%
\providecommand \enquote  [1]{``#1''}%
\providecommand \bibnamefont  [1]{#1}%
\providecommand \bibfnamefont [1]{#1}%
\providecommand \citenamefont [1]{#1}%
\providecommand \href@noop [0]{\@secondoftwo}%
\providecommand \href [0]{\begingroup \@sanitize@url \@href}%
\providecommand \@href[1]{\@@startlink{#1}\@@href}%
\providecommand \@@href[1]{\endgroup#1\@@endlink}%
\providecommand \@sanitize@url [0]{\catcode `\\12\catcode `\$12\catcode
  `\&12\catcode `\#12\catcode `\^12\catcode `\_12\catcode `\%12\relax}%
\providecommand \@@startlink[1]{}%
\providecommand \@@endlink[0]{}%
\providecommand \url  [0]{\begingroup\@sanitize@url \@url }%
\providecommand \@url [1]{\endgroup\@href {#1}{\urlprefix }}%
\providecommand \urlprefix  [0]{URL }%
\providecommand \Eprint [0]{\href }%
\providecommand \doibase [0]{http://dx.doi.org/}%
\providecommand \selectlanguage [0]{\@gobble}%
\providecommand \bibinfo  [0]{\@secondoftwo}%
\providecommand \bibfield  [0]{\@secondoftwo}%
\providecommand \translation [1]{[#1]}%
\providecommand \BibitemOpen [0]{}%
\providecommand \bibitemStop [0]{}%
\providecommand \bibitemNoStop [0]{.\EOS\space}%
\providecommand \EOS [0]{\spacefactor3000\relax}%
\providecommand \BibitemShut  [1]{\csname bibitem#1\endcsname}%
\let\auto@bib@innerbib\@empty
\bibitem [{\citenamefont {Schill}\ \emph {et~al.}(2024)\citenamefont {Schill},
  \citenamefont {Armstrong}, \citenamefont {Nguyen}, \citenamefont {Sterbentz},
  \citenamefont {White}, \citenamefont {Benedict}, \citenamefont {Rieben},
  \citenamefont {Hoff}, \citenamefont {Lorenzana}, \citenamefont {Belof} \emph
  {et~al.}}]{schill2024suppression}%
  \BibitemOpen
  \bibfield  {author} {\bibinfo {author} {\bibfnamefont {W.~J.}\ \bibnamefont
  {Schill}}, \bibinfo {author} {\bibfnamefont {M.~R.}\ \bibnamefont
  {Armstrong}}, \bibinfo {author} {\bibfnamefont {J.~H.}\ \bibnamefont
  {Nguyen}}, \bibinfo {author} {\bibfnamefont {D.~M.}\ \bibnamefont
  {Sterbentz}}, \bibinfo {author} {\bibfnamefont {D.~A.}\ \bibnamefont
  {White}}, \bibinfo {author} {\bibfnamefont {L.~X.}\ \bibnamefont {Benedict}},
  \bibinfo {author} {\bibfnamefont {R.~N.}\ \bibnamefont {Rieben}}, \bibinfo
  {author} {\bibfnamefont {A.}~\bibnamefont {Hoff}}, \bibinfo {author}
  {\bibfnamefont {H.~E.}\ \bibnamefont {Lorenzana}}, \bibinfo {author}
  {\bibfnamefont {J.~L.}\ \bibnamefont {Belof}},  \emph {et~al.},\ }\bibfield
  {title} {\enquote {\bibinfo {title} {Suppression of richtmyer-meshkov
  instability via special pairs of shocks and phase transitions},}\ }\href
  {\doibase {}} {\bibfield  {journal} {\bibinfo  {journal} {Phys. Rev. Lett.}\
  }\textbf {\bibinfo {volume} {132}},\ \bibinfo {pages} {024001} (\bibinfo
  {year} {2024})}\BibitemShut {NoStop}%
\bibitem [{\citenamefont {Zhou}, \citenamefont {Sadler},\ and\ \citenamefont
  {Hurricane}(2025)}]{zhou2025instabilities}%
  \BibitemOpen
  \bibfield  {author} {\bibinfo {author} {\bibfnamefont {Y.}~\bibnamefont
  {Zhou}}, \bibinfo {author} {\bibfnamefont {J.~D.}\ \bibnamefont {Sadler}}, \
  and\ \bibinfo {author} {\bibfnamefont {O.~A.}\ \bibnamefont {Hurricane}},\
  }\bibfield  {title} {\enquote {\bibinfo {title} {Instabilities and mixing in
  inertial confinement fusion},}\ }\href {\doibase {}} {\bibfield  {journal}
  {\bibinfo  {journal} {Annu. Rev. Fluid Mech.}\ }\textbf {\bibinfo {volume}
  {57}},\ \bibinfo {pages} {197--225} (\bibinfo {year} {2025})}\BibitemShut
  {NoStop}%
\bibitem [{\citenamefont {Abarzhi}(2024)}]{abarzhi2024perspective}%
  \BibitemOpen
  \bibfield  {author} {\bibinfo {author} {\bibfnamefont {S.~I.}\ \bibnamefont
  {Abarzhi}},\ }\bibfield  {title} {\enquote {\bibinfo {title} {{Perspective:
  group theory analysis and special self-similarity classes in Rayleigh--Taylor
  and Richtmyer--Meshkov interfacial mixing with variable accelerations}},}\
  }\href {\doibase {https://doi.org/10.1007/s41614-023-00142-3}} {\bibfield
  {journal} {\bibinfo  {journal} {Rev. Mod. Plasma Phys.}\ }\textbf {\bibinfo
  {volume} {8}},\ \bibinfo {pages} {15} (\bibinfo {year} {2024})}\BibitemShut
  {NoStop}%
\bibitem [{\citenamefont {Zhou}(2024)}]{zhou2024hydrodynamic}%
  \BibitemOpen
  \bibfield  {author} {\bibinfo {author} {\bibfnamefont {Y.}~\bibnamefont
  {Zhou}},\ }\href {\doibase {}} {\emph {\bibinfo {title} {{Hydrodynamic
  Instabilities and Turbulence: Rayleigh--Taylor, Richtmyer--Meshkov, and
  Kelvin--Helmholtz Mixing}}}}\ (\bibinfo  {publisher} {Cambridge University
  Press},\ \bibinfo {address} {Cambridge, UK},\ \bibinfo {year}
  {2024})\BibitemShut {NoStop}%
\bibitem [{\citenamefont {Mostert}\ \emph {et~al.}(2015)\citenamefont
  {Mostert}, \citenamefont {Wheatley}, \citenamefont {Samtaney},\ and\
  \citenamefont {Pullin}}]{mostert2015effects}%
  \BibitemOpen
  \bibfield  {author} {\bibinfo {author} {\bibfnamefont {W.}~\bibnamefont
  {Mostert}}, \bibinfo {author} {\bibfnamefont {V.}~\bibnamefont {Wheatley}},
  \bibinfo {author} {\bibfnamefont {R.}~\bibnamefont {Samtaney}}, \ and\
  \bibinfo {author} {\bibfnamefont {D.~I.}\ \bibnamefont {Pullin}},\ }\bibfield
   {title} {\enquote {\bibinfo {title} {{Effects of magnetic fields on
  magnetohydrodynamic cylindrical and spherical Richtmyer--Meshkov
  instability}},}\ }\href {\doibase {}} {\bibfield  {journal} {\bibinfo
  {journal} {Phys. Fluids}\ }\textbf {\bibinfo {volume} {27}},\ \bibinfo
  {pages} {104102} (\bibinfo {year} {2015})}\BibitemShut {NoStop}%
\bibitem [{\citenamefont {Attal}\ and\ \citenamefont
  {Ramaprabhu}(2015)}]{attal2015numerical}%
  \BibitemOpen
  \bibfield  {author} {\bibinfo {author} {\bibfnamefont {N.}~\bibnamefont
  {Attal}}\ and\ \bibinfo {author} {\bibfnamefont {P.}~\bibnamefont
  {Ramaprabhu}},\ }\bibfield  {title} {\enquote {\bibinfo {title} {{Numerical
  investigation of a single-mode chemically reacting Richtmyer--Meshkov
  instability}},}\ }\href {\doibase {}} {\bibfield  {journal} {\bibinfo
  {journal} {Shock Waves}\ }\textbf {\bibinfo {volume} {25}},\ \bibinfo {pages}
  {307--328} (\bibinfo {year} {2015})}\BibitemShut {NoStop}%
\bibitem [{\citenamefont {Richtmyer}(1960)}]{1960Richtmyer}%
  \BibitemOpen
  \bibfield  {author} {\bibinfo {author} {\bibfnamefont {R.~D.}\ \bibnamefont
  {Richtmyer}},\ }\bibfield  {title} {\enquote {\bibinfo {title} {Taylor
  instability in shock acceleration of compressible fluids},}\ }\href {\doibase
  {}} {\bibfield  {journal} {\bibinfo  {journal} {Commun. Pur. Appl. Math.}\
  }\textbf {\bibinfo {volume} {13}},\ \bibinfo {pages} {297--319} (\bibinfo
  {year} {1960})}\BibitemShut {NoStop}%
\bibitem [{\citenamefont {Meshkov}(1969)}]{meshkov1969instability}%
  \BibitemOpen
  \bibfield  {author} {\bibinfo {author} {\bibfnamefont {E.~E.}\ \bibnamefont
  {Meshkov}},\ }\bibfield  {title} {\enquote {\bibinfo {title} {Instability of
  the interface of two gases accelerated by a shock wave},}\ }\href {\doibase
  {}} {\bibfield  {journal} {\bibinfo  {journal} {Fluid Dynam.}\ }\textbf
  {\bibinfo {volume} {4}},\ \bibinfo {pages} {101--104} (\bibinfo {year}
  {1969})}\BibitemShut {NoStop}%
\bibitem [{\citenamefont {Xu}\ \emph {et~al.}(2022)\citenamefont {Xu},
  \citenamefont {Cong}, \citenamefont {Si},\ and\ \citenamefont
  {Luo}}]{guo2022shock}%
  \BibitemOpen
  \bibfield  {author} {\bibinfo {author} {\bibfnamefont {A.~G.}\ \bibnamefont
  {Xu}}, \bibinfo {author} {\bibfnamefont {Z.~Y.}\ \bibnamefont {Cong}},
  \bibinfo {author} {\bibfnamefont {T.}~\bibnamefont {Si}}, \ and\ \bibinfo
  {author} {\bibfnamefont {X.~S.}\ \bibnamefont {Luo}},\ }\bibfield  {title}
  {\enquote {\bibinfo {title} {{Shock-tube studies of single-and
  quasi-single-mode perturbation growth in Richtmyer--Meshkov flows with
  reshock}},}\ }\href {\doibase {}} {\bibfield  {journal} {\bibinfo  {journal}
  {J. Fluid Mech.}\ }\textbf {\bibinfo {volume} {941}},\ \bibinfo {pages} {A65}
  (\bibinfo {year} {2022})}\BibitemShut {NoStop}%
\bibitem [{\citenamefont {Wang}\ \emph {et~al.}(2023)\citenamefont {Wang},
  \citenamefont {Wang}, \citenamefont {Zhai},\ and\ \citenamefont
  {Luo}}]{wang2023high}%
  \BibitemOpen
  \bibfield  {author} {\bibinfo {author} {\bibfnamefont {H.}~\bibnamefont
  {Wang}}, \bibinfo {author} {\bibfnamefont {H.}~\bibnamefont {Wang}}, \bibinfo
  {author} {\bibfnamefont {Z.~G.}\ \bibnamefont {Zhai}}, \ and\ \bibinfo
  {author} {\bibfnamefont {X.~S.}\ \bibnamefont {Luo}},\ }\bibfield  {title}
  {\enquote {\bibinfo {title} {{High-amplitude effect on single-mode
  Richtmyer--Meshkov instability of a light-heavy interface}},}\ }\href
  {\doibase {}} {\bibfield  {journal} {\bibinfo  {journal} {Phys. Fluids}\
  }\textbf {\bibinfo {volume} {35}},\ \bibinfo {pages} {126107} (\bibinfo
  {year} {2023})}\BibitemShut {NoStop}%
\bibitem [{\citenamefont {Latini}, \citenamefont {Schilling},\ and\
  \citenamefont {Meiron}(2024)}]{latini2024analysis}%
  \BibitemOpen
  \bibfield  {author} {\bibinfo {author} {\bibfnamefont {M.}~\bibnamefont
  {Latini}}, \bibinfo {author} {\bibfnamefont {O.}~\bibnamefont {Schilling}}, \
  and\ \bibinfo {author} {\bibfnamefont {D.~I.}\ \bibnamefont {Meiron}},\
  }\bibfield  {title} {\enquote {\bibinfo {title} {{Analysis of single-mode
  Richtmyer--Meshkov instability using high-order incompressible
  vorticity-streamfunction and shock-capturing simulations}},}\ }\href
  {\doibase {}} {\bibfield  {journal} {\bibinfo  {journal} {Phys. Fluids}\
  }\textbf {\bibinfo {volume} {36}},\ \bibinfo {pages} {024115} (\bibinfo
  {year} {2024})}\BibitemShut {NoStop}%
\bibitem [{\citenamefont {Brouillette}\ and\ \citenamefont
  {Sturtevant}(1989)}]{brouillette1989growth}%
  \BibitemOpen
  \bibfield  {author} {\bibinfo {author} {\bibfnamefont {M.}~\bibnamefont
  {Brouillette}}\ and\ \bibinfo {author} {\bibfnamefont {B.}~\bibnamefont
  {Sturtevant}},\ }\bibfield  {title} {\enquote {\bibinfo {title} {Growth
  induced by multiple shock waves normally incident on plane gaseous
  interfaces},}\ }\href {\doibase {}} {\bibfield  {journal} {\bibinfo
  {journal} {Physica D}\ }\textbf {\bibinfo {volume} {37}},\ \bibinfo {pages}
  {248--263} (\bibinfo {year} {1989})}\BibitemShut {NoStop}%
\bibitem [{\citenamefont {Charakhch'yan}(2001)}]{charakhch2001reshocking}%
  \BibitemOpen
  \bibfield  {author} {\bibinfo {author} {\bibfnamefont {A.~A.}\ \bibnamefont
  {Charakhch'yan}},\ }\bibfield  {title} {\enquote {\bibinfo {title}
  {{Reshocking at the non-linear stage of Richtmyer--Meshkov instability}},}\
  }\href {\doibase {}} {\bibfield  {journal} {\bibinfo  {journal} {Plasma Phys.
  Control. Fusion}\ }\textbf {\bibinfo {volume} {43}},\ \bibinfo {pages} {1169}
  (\bibinfo {year} {2001})}\BibitemShut {NoStop}%
\bibitem [{\citenamefont {Mikaelian}(1989)}]{mikaelian1989turbulent}%
  \BibitemOpen
  \bibfield  {author} {\bibinfo {author} {\bibfnamefont {K.~O.}\ \bibnamefont
  {Mikaelian}},\ }\bibfield  {title} {\enquote {\bibinfo {title} {{Turbulent
  mixing generated by Rayleigh--Taylor and Richtmyer--Meshkov
  instabilities}},}\ }\href {\doibase {}} {\bibfield  {journal} {\bibinfo
  {journal} {Physica D}\ }\textbf {\bibinfo {volume} {36}},\ \bibinfo {pages}
  {343--357} (\bibinfo {year} {1989})}\BibitemShut {NoStop}%
\bibitem [{\citenamefont {Mikaelian}(2011)}]{mikaelian2011extended}%
  \BibitemOpen
  \bibfield  {author} {\bibinfo {author} {\bibfnamefont {K.~O.}\ \bibnamefont
  {Mikaelian}},\ }\bibfield  {title} {\enquote {\bibinfo {title} {{Extended
  model for Richtmyer--Meshkov mix}},}\ }\href {\doibase {}} {\bibfield
  {journal} {\bibinfo  {journal} {Physica D}\ }\textbf {\bibinfo {volume}
  {240}},\ \bibinfo {pages} {935--942} (\bibinfo {year} {2011})}\BibitemShut
  {NoStop}%
\bibitem [{\citenamefont {Mikaelian}(2015)}]{mikaelian2015testing}%
  \BibitemOpen
  \bibfield  {author} {\bibinfo {author} {\bibfnamefont {K.~O.}\ \bibnamefont
  {Mikaelian}},\ }\bibfield  {title} {\enquote {\bibinfo {title} {{Testing an
  analytic model for Richtmyer--Meshkov turbulent mixing widths}},}\ }\href
  {\doibase {}} {\bibfield  {journal} {\bibinfo  {journal} {Shock Waves}\
  }\textbf {\bibinfo {volume} {25}},\ \bibinfo {pages} {35--45} (\bibinfo
  {year} {2015})}\BibitemShut {NoStop}%
\bibitem [{\citenamefont {Zhang}\ \emph {et~al.}(2023)\citenamefont {Zhang},
  \citenamefont {Zhao}, \citenamefont {Ding},\ and\ \citenamefont
  {Luo}}]{zhang2023richtmyer}%
  \BibitemOpen
  \bibfield  {author} {\bibinfo {author} {\bibfnamefont {Y.~M.}\ \bibnamefont
  {Zhang}}, \bibinfo {author} {\bibfnamefont {Y.}~\bibnamefont {Zhao}},
  \bibinfo {author} {\bibfnamefont {J.~C.}\ \bibnamefont {Ding}}, \ and\
  \bibinfo {author} {\bibfnamefont {X.~S.}\ \bibnamefont {Luo}},\ }\bibfield
  {title} {\enquote {\bibinfo {title} {{Richtmyer--Meshkov instability with a
  rippled reshock}},}\ }\href {\doibase {}} {\bibfield  {journal} {\bibinfo
  {journal} {J. Fluid Mech.}\ }\textbf {\bibinfo {volume} {968}},\ \bibinfo
  {pages} {A3} (\bibinfo {year} {2023})}\BibitemShut {NoStop}%
\bibitem [{\citenamefont {Hosseini}\ and\ \citenamefont
  {Takayama}(2005)}]{hosseini2005experimental}%
  \BibitemOpen
  \bibfield  {author} {\bibinfo {author} {\bibfnamefont {S.~H.~R.}\
  \bibnamefont {Hosseini}}\ and\ \bibinfo {author} {\bibfnamefont
  {K.}~\bibnamefont {Takayama}},\ }\bibfield  {title} {\enquote {\bibinfo
  {title} {{Experimental study of Richtmyer--Meshkov instability induced by
  cylindrical shock waves}},}\ }\href {\doibase {}} {\bibfield  {journal}
  {\bibinfo  {journal} {Phys. Fluids}\ }\textbf {\bibinfo {volume} {17}},\
  \bibinfo {pages} {084101} (\bibinfo {year} {2005})}\BibitemShut {NoStop}%
\bibitem [{\citenamefont {Jacobs}\ \emph {et~al.}(2013)\citenamefont {Jacobs},
  \citenamefont {Krivets}, \citenamefont {Tsiklashvili},\ and\ \citenamefont
  {Likhachev}}]{jacobs2013experiments}%
  \BibitemOpen
  \bibfield  {author} {\bibinfo {author} {\bibfnamefont {J.~W.}\ \bibnamefont
  {Jacobs}}, \bibinfo {author} {\bibfnamefont {V.~V.}\ \bibnamefont {Krivets}},
  \bibinfo {author} {\bibfnamefont {V.}~\bibnamefont {Tsiklashvili}}, \ and\
  \bibinfo {author} {\bibfnamefont {O.~A.}\ \bibnamefont {Likhachev}},\
  }\bibfield  {title} {\enquote {\bibinfo {title} {{Experiments on the
  Richtmyer--Meshkov instability with an imposed, random initial
  perturbation}},}\ }\href {\doibase {}} {\bibfield  {journal} {\bibinfo
  {journal} {Shock Waves}\ }\textbf {\bibinfo {volume} {23}},\ \bibinfo {pages}
  {407--413} (\bibinfo {year} {2013})}\BibitemShut {NoStop}%
\bibitem [{\citenamefont {Guo}\ \emph {et~al.}(2024)\citenamefont {Guo},
  \citenamefont {Cong}, \citenamefont {Si},\ and\ \citenamefont
  {Luo}}]{guo2024richtmyer}%
  \BibitemOpen
  \bibfield  {author} {\bibinfo {author} {\bibfnamefont {X.}~\bibnamefont
  {Guo}}, \bibinfo {author} {\bibfnamefont {Z.~Y.}\ \bibnamefont {Cong}},
  \bibinfo {author} {\bibfnamefont {T.}~\bibnamefont {Si}}, \ and\ \bibinfo
  {author} {\bibfnamefont {X.~S.}\ \bibnamefont {Luo}},\ }\bibfield  {title}
  {\enquote {\bibinfo {title} {{On Richtmyer--Meshkov finger collisions in a
  light fluid layer under reshock conditions}},}\ }\href {\doibase {}}
  {\bibfield  {journal} {\bibinfo  {journal} {J. Fluid Mech.}\ }\textbf
  {\bibinfo {volume} {1000}},\ \bibinfo {pages} {A87} (\bibinfo {year}
  {2024})}\BibitemShut {NoStop}%
\bibitem [{\citenamefont {Zhai}\ \emph {et~al.}(2018)\citenamefont {Zhai},
  \citenamefont {Zou}, \citenamefont {Wu},\ and\ \citenamefont
  {Luo}}]{zhai2018review}%
  \BibitemOpen
  \bibfield  {author} {\bibinfo {author} {\bibfnamefont {Z.~G.}\ \bibnamefont
  {Zhai}}, \bibinfo {author} {\bibfnamefont {L.~Y.}\ \bibnamefont {Zou}},
  \bibinfo {author} {\bibfnamefont {Q.}~\bibnamefont {Wu}}, \ and\ \bibinfo
  {author} {\bibfnamefont {X.~S.}\ \bibnamefont {Luo}},\ }\bibfield  {title}
  {\enquote {\bibinfo {title} {{Review of experimental Richtmyer--Meshkov
  instability in shock tube: from simple to complex}},}\ }\href {\doibase {}}
  {\bibfield  {journal} {\bibinfo  {journal} {Proc. Inst. Mech. Eng. C J. Mech.
  Eng. Sci.}\ }\textbf {\bibinfo {volume} {232}},\ \bibinfo {pages}
  {2830--2849} (\bibinfo {year} {2018})}\BibitemShut {NoStop}%
\bibitem [{\citenamefont {Chen}\ \emph
  {et~al.}(2024{\natexlab{a}})\citenamefont {Chen}, \citenamefont {Jin},
  \citenamefont {Liang},\ and\ \citenamefont {Zou}}]{chen2024numerical}%
  \BibitemOpen
  \bibfield  {author} {\bibinfo {author} {\bibfnamefont {Y.~F.}\ \bibnamefont
  {Chen}}, \bibinfo {author} {\bibfnamefont {T.}~\bibnamefont {Jin}}, \bibinfo
  {author} {\bibfnamefont {Z.~H.}\ \bibnamefont {Liang}}, \ and\ \bibinfo
  {author} {\bibfnamefont {L.~Y.}\ \bibnamefont {Zou}},\ }\bibfield  {title}
  {\enquote {\bibinfo {title} {{Numerical study of shock-induced
  Richtmyer--Meshkov instability in inhomogeneous heavy fluid layer}},}\ }\href
  {\doibase {}} {\bibfield  {journal} {\bibinfo  {journal} {Phys. Fluids}\
  }\textbf {\bibinfo {volume} {36}},\ \bibinfo {pages} {092105} (\bibinfo
  {year} {2024}{\natexlab{a}})}\BibitemShut {NoStop}%
\bibitem [{\citenamefont {Leinov}\ \emph {et~al.}(2009)\citenamefont {Leinov},
  \citenamefont {Malamud}, \citenamefont {Elbaz}, \citenamefont {Levin},
  \citenamefont {Ben-Dor}, \citenamefont {Shvarts},\ and\ \citenamefont
  {Sadot}}]{leinov2009experimental}%
  \BibitemOpen
  \bibfield  {author} {\bibinfo {author} {\bibfnamefont {E.}~\bibnamefont
  {Leinov}}, \bibinfo {author} {\bibfnamefont {G.}~\bibnamefont {Malamud}},
  \bibinfo {author} {\bibfnamefont {Y.}~\bibnamefont {Elbaz}}, \bibinfo
  {author} {\bibfnamefont {L.}~\bibnamefont {Levin}}, \bibinfo {author}
  {\bibfnamefont {G.}~\bibnamefont {Ben-Dor}}, \bibinfo {author} {\bibfnamefont
  {D.}~\bibnamefont {Shvarts}}, \ and\ \bibinfo {author} {\bibfnamefont
  {O.}~\bibnamefont {Sadot}},\ }\bibfield  {title} {\enquote {\bibinfo {title}
  {{Experimental and numerical investigation of the Richtmyer--Meshkov
  instability under re-shock conditions}},}\ }\href {\doibase {}} {\bibfield
  {journal} {\bibinfo  {journal} {J. Fluid Mech.}\ }\textbf {\bibinfo {volume}
  {626}},\ \bibinfo {pages} {449--475} (\bibinfo {year} {2009})}\BibitemShut
  {NoStop}%
\bibitem [{\citenamefont {Si}\ \emph {et~al.}(2012)\citenamefont {Si},
  \citenamefont {Zhai}, \citenamefont {Yang},\ and\ \citenamefont
  {Luo}}]{si2012experimental}%
  \BibitemOpen
  \bibfield  {author} {\bibinfo {author} {\bibfnamefont {T.}~\bibnamefont
  {Si}}, \bibinfo {author} {\bibfnamefont {Z.~G.}\ \bibnamefont {Zhai}},
  \bibinfo {author} {\bibfnamefont {J.~M.}\ \bibnamefont {Yang}}, \ and\
  \bibinfo {author} {\bibfnamefont {X.~S.}\ \bibnamefont {Luo}},\ }\bibfield
  {title} {\enquote {\bibinfo {title} {Experimental investigation of reshocked
  spherical gas interfaces},}\ }\href {\doibase {}} {\bibfield  {journal}
  {\bibinfo  {journal} {Phys. Fluids}\ }\textbf {\bibinfo {volume} {24}},\
  \bibinfo {pages} {054101} (\bibinfo {year} {2012})}\BibitemShut {NoStop}%
\bibitem [{\citenamefont {Ding}, \citenamefont {Deng},\ and\ \citenamefont
  {Luo}(2021)}]{ding2021convergent}%
  \BibitemOpen
  \bibfield  {author} {\bibinfo {author} {\bibfnamefont {J.~C.}\ \bibnamefont
  {Ding}}, \bibinfo {author} {\bibfnamefont {X.~M.}\ \bibnamefont {Deng}}, \
  and\ \bibinfo {author} {\bibfnamefont {X.~S.}\ \bibnamefont {Luo}},\
  }\bibfield  {title} {\enquote {\bibinfo {title} {{Convergent
  Richtmyer--Meshkov instability on a light gas layer with perturbed inner and
  outer surfaces}},}\ }\href {\doibase {}} {\bibfield  {journal} {\bibinfo
  {journal} {Phys. Fluids}\ }\textbf {\bibinfo {volume} {33}},\ \bibinfo
  {pages} {102112} (\bibinfo {year} {2021})}\BibitemShut {NoStop}%
\bibitem [{\citenamefont {Ferguson}\ and\ \citenamefont
  {Jacobs}(2024)}]{ferguson2024influence}%
  \BibitemOpen
  \bibfield  {author} {\bibinfo {author} {\bibfnamefont {K.}~\bibnamefont
  {Ferguson}}\ and\ \bibinfo {author} {\bibfnamefont {J.~W.}\ \bibnamefont
  {Jacobs}},\ }\bibfield  {title} {\enquote {\bibinfo {title} {{The influence
  of the shock-to-reshock time on the Richtmyer--Meshkov instability in
  reshock}},}\ }\href {\doibase {}} {\bibfield  {journal} {\bibinfo  {journal}
  {J. Fluid Mech.}\ }\textbf {\bibinfo {volume} {999}},\ \bibinfo {pages} {A68}
  (\bibinfo {year} {2024})}\BibitemShut {NoStop}%
\bibitem [{\citenamefont {Chen}\ \emph {et~al.}(2023)\citenamefont {Chen},
  \citenamefont {Jin}, \citenamefont {Liang},\ and\ \citenamefont
  {Zou}}]{chen2023numerical}%
  \BibitemOpen
  \bibfield  {author} {\bibinfo {author} {\bibfnamefont {Y.~F.}\ \bibnamefont
  {Chen}}, \bibinfo {author} {\bibfnamefont {T.}~\bibnamefont {Jin}}, \bibinfo
  {author} {\bibfnamefont {Z.~G.}\ \bibnamefont {Liang}}, \ and\ \bibinfo
  {author} {\bibfnamefont {L.~Y.}\ \bibnamefont {Zou}},\ }\bibfield  {title}
  {\enquote {\bibinfo {title} {{Numerical study of Richtmyer--Meshkov
  instability of light fluid layer with reshock}},}\ }\href {\doibase {}}
  {\bibfield  {journal} {\bibinfo  {journal} {Phys. Fluids}\ }\textbf {\bibinfo
  {volume} {35}},\ \bibinfo {pages} {114103} (\bibinfo {year}
  {2023})}\BibitemShut {NoStop}%
\bibitem [{\citenamefont {Wang}\ \emph {et~al.}(2010)\citenamefont {Wang},
  \citenamefont {Bai}, \citenamefont {Li},\ and\ \citenamefont
  {Liu}}]{wang2010large}%
  \BibitemOpen
  \bibfield  {author} {\bibinfo {author} {\bibfnamefont {T.}~\bibnamefont
  {Wang}}, \bibinfo {author} {\bibfnamefont {J.~S.}\ \bibnamefont {Bai}},
  \bibinfo {author} {\bibfnamefont {P.}~\bibnamefont {Li}}, \ and\ \bibinfo
  {author} {\bibfnamefont {K.}~\bibnamefont {Liu}},\ }\bibfield  {title}
  {\enquote {\bibinfo {title} {{Large-eddy simulations of the Richtmyer-Meshkov
  instability of rectangular interfaces accelerated by shock waves}},}\ }\href
  {\doibase {}} {\bibfield  {journal} {\bibinfo  {journal} {Sci. China Phys.
  Mech. Astron.}\ }\textbf {\bibinfo {volume} {53}},\ \bibinfo {pages}
  {905--914} (\bibinfo {year} {2010})}\BibitemShut {NoStop}%
\bibitem [{\citenamefont {Lombardini}\ \emph {et~al.}(2011)\citenamefont
  {Lombardini}, \citenamefont {Hill}, \citenamefont {Pullin},\ and\
  \citenamefont {Meiron}}]{lombardini2011atwood}%
  \BibitemOpen
  \bibfield  {author} {\bibinfo {author} {\bibfnamefont {M.}~\bibnamefont
  {Lombardini}}, \bibinfo {author} {\bibfnamefont {D.}~\bibnamefont {Hill}},
  \bibinfo {author} {\bibfnamefont {D.}~\bibnamefont {Pullin}}, \ and\ \bibinfo
  {author} {\bibfnamefont {D.}~\bibnamefont {Meiron}},\ }\bibfield  {title}
  {\enquote {\bibinfo {title} {{Atwood ratio dependence of Richtmyer--Meshkov
  flows under reshock conditions using large-eddy simulations}},}\ }\href
  {\doibase {}} {\bibfield  {journal} {\bibinfo  {journal} {J. Fluid Mech.}\
  }\textbf {\bibinfo {volume} {670}},\ \bibinfo {pages} {439--480} (\bibinfo
  {year} {2011})}\BibitemShut {NoStop}%
\bibitem [{\citenamefont {Bin}\ \emph {et~al.}(2021)\citenamefont {Bin},
  \citenamefont {Xiao}, \citenamefont {Shi}, \citenamefont {Zhang},\ and\
  \citenamefont {Chen}}]{bin2021new}%
  \BibitemOpen
  \bibfield  {author} {\bibinfo {author} {\bibfnamefont {Y.~W.}\ \bibnamefont
  {Bin}}, \bibinfo {author} {\bibfnamefont {M.~J.}\ \bibnamefont {Xiao}},
  \bibinfo {author} {\bibfnamefont {Y.~P.}\ \bibnamefont {Shi}}, \bibinfo
  {author} {\bibfnamefont {Y.~S.}\ \bibnamefont {Zhang}}, \ and\ \bibinfo
  {author} {\bibfnamefont {S.~Y.}\ \bibnamefont {Chen}},\ }\bibfield  {title}
  {\enquote {\bibinfo {title} {{A new idea to predict reshocked
  Richtmyer--Meshkov mixing: Constrained large-eddy simulation}},}\ }\href
  {\doibase {}} {\bibfield  {journal} {\bibinfo  {journal} {J. Fluid Mech.}\
  }\textbf {\bibinfo {volume} {918}},\ \bibinfo {pages} {R1} (\bibinfo {year}
  {2021})}\BibitemShut {NoStop}%
\bibitem [{\citenamefont {Sadler}\ \emph {et~al.}(2024)\citenamefont {Sadler},
  \citenamefont {Powell}, \citenamefont {Schalles}, \citenamefont {Louie},
  \citenamefont {Jacobs},\ and\ \citenamefont {Zhou}}]{sadler2024simulations}%
  \BibitemOpen
  \bibfield  {author} {\bibinfo {author} {\bibfnamefont {J.~D.}\ \bibnamefont
  {Sadler}}, \bibinfo {author} {\bibfnamefont {P.~D.}\ \bibnamefont {Powell}},
  \bibinfo {author} {\bibfnamefont {M.}~\bibnamefont {Schalles}}, \bibinfo
  {author} {\bibfnamefont {C.}~\bibnamefont {Louie}}, \bibinfo {author}
  {\bibfnamefont {J.~W.}\ \bibnamefont {Jacobs}}, \ and\ \bibinfo {author}
  {\bibfnamefont {Y.}~\bibnamefont {Zhou}},\ }\bibfield  {title} {\enquote
  {\bibinfo {title} {{Simulations of three-layer Richtmyer--Meshkov mixing in a
  shock tube}},}\ }\href {\doibase {}} {\bibfield  {journal} {\bibinfo
  {journal} {Phys. Fluids}\ }\textbf {\bibinfo {volume} {36}},\ \bibinfo
  {pages} {014120} (\bibinfo {year} {2024})}\BibitemShut {NoStop}%
\bibitem [{\citenamefont {Xu}, \citenamefont {Zhang},\ and\ \citenamefont
  {Gan}(2024)}]{xu2024advances}%
  \BibitemOpen
  \bibfield  {author} {\bibinfo {author} {\bibfnamefont {A.~G.}\ \bibnamefont
  {Xu}}, \bibinfo {author} {\bibfnamefont {D.~J.}\ \bibnamefont {Zhang}}, \
  and\ \bibinfo {author} {\bibfnamefont {Y.~B.}\ \bibnamefont {Gan}},\
  }\bibfield  {title} {\enquote {\bibinfo {title} {Advances in the kinetics of
  heat and mass transfer in near-continuous complex flows},}\ }\href {\doibase
  {}} {\bibfield  {journal} {\bibinfo  {journal} {Front. Phys.}\ }\textbf
  {\bibinfo {volume} {19}},\ \bibinfo {pages} {42500} (\bibinfo {year}
  {2024})}\BibitemShut {NoStop}%
\bibitem [{\citenamefont {Gan}\ \emph {et~al.}(2018)\citenamefont {Gan},
  \citenamefont {Xu}, \citenamefont {Zhang}, \citenamefont {Zhang},\ and\
  \citenamefont {Succi}}]{gan2018discrete}%
  \BibitemOpen
  \bibfield  {author} {\bibinfo {author} {\bibfnamefont {Y.~B.}\ \bibnamefont
  {Gan}}, \bibinfo {author} {\bibfnamefont {A.~G.}\ \bibnamefont {Xu}},
  \bibinfo {author} {\bibfnamefont {G.~C.}\ \bibnamefont {Zhang}}, \bibinfo
  {author} {\bibfnamefont {Y.~D.}\ \bibnamefont {Zhang}}, \ and\ \bibinfo
  {author} {\bibfnamefont {S.}~\bibnamefont {Succi}},\ }\bibfield  {title}
  {\enquote {\bibinfo {title} {{Discrete Boltzmann trans-scale modeling of
  high-speed compressible flows}},}\ }\href {\doibase {}} {\bibfield  {journal}
  {\bibinfo  {journal} {Phys. Rev. E}\ }\textbf {\bibinfo {volume} {97}},\
  \bibinfo {pages} {053312} (\bibinfo {year} {2018})}\BibitemShut {NoStop}%
\bibitem [{\citenamefont {Lin}\ \emph {et~al.}(2017)\citenamefont {Lin},
  \citenamefont {Luo}, \citenamefont {Fei},\ and\ \citenamefont
  {Succi}}]{lin2017multi}%
  \BibitemOpen
  \bibfield  {author} {\bibinfo {author} {\bibfnamefont {C.~D.}\ \bibnamefont
  {Lin}}, \bibinfo {author} {\bibfnamefont {K.~H.}\ \bibnamefont {Luo}},
  \bibinfo {author} {\bibfnamefont {L.~L.}\ \bibnamefont {Fei}}, \ and\
  \bibinfo {author} {\bibfnamefont {S.}~\bibnamefont {Succi}},\ }\bibfield
  {title} {\enquote {\bibinfo {title} {{A multi-component discrete Boltzmann
  model for nonequilibrium reactive flows}},}\ }\href {\doibase {}} {\bibfield
  {journal} {\bibinfo  {journal} {Sci. Rep.}\ }\textbf {\bibinfo {volume}
  {7}},\ \bibinfo {pages} {14580} (\bibinfo {year} {2017})}\BibitemShut
  {NoStop}%
\bibitem [{\citenamefont {Lin}\ and\ \citenamefont
  {Luo}(2020)}]{lin2020kinetic}%
  \BibitemOpen
  \bibfield  {author} {\bibinfo {author} {\bibfnamefont {C.~D.}\ \bibnamefont
  {Lin}}\ and\ \bibinfo {author} {\bibfnamefont {K.~H.}\ \bibnamefont {Luo}},\
  }\bibfield  {title} {\enquote {\bibinfo {title} {Kinetic simulation of
  unsteady detonation with thermodynamic nonequilibrium effects},}\ }\href
  {\doibase {}} {\bibfield  {journal} {\bibinfo  {journal} {Combust. Explos.
  Shock Waves}\ }\textbf {\bibinfo {volume} {56}},\ \bibinfo {pages} {435--443}
  (\bibinfo {year} {2020})}\BibitemShut {NoStop}%
\bibitem [{\citenamefont {Ji}, \citenamefont {Lin},\ and\ \citenamefont
  {Luo}(2022)}]{ji2022three}%
  \BibitemOpen
  \bibfield  {author} {\bibinfo {author} {\bibfnamefont {Y.}~\bibnamefont
  {Ji}}, \bibinfo {author} {\bibfnamefont {C.~D.}\ \bibnamefont {Lin}}, \ and\
  \bibinfo {author} {\bibfnamefont {K.~H.}\ \bibnamefont {Luo}},\ }\bibfield
  {title} {\enquote {\bibinfo {title} {{A three-dimensional discrete Boltzmann
  model for steady and unsteady detonation}},}\ }\href {\doibase {}} {\bibfield
   {journal} {\bibinfo  {journal} {J. Comput. Phys.}\ }\textbf {\bibinfo
  {volume} {455}},\ \bibinfo {pages} {111002} (\bibinfo {year}
  {2022})}\BibitemShut {NoStop}%
\bibitem [{\citenamefont {Su}\ and\ \citenamefont
  {Lin}(2023)}]{su2023unsteady}%
  \BibitemOpen
  \bibfield  {author} {\bibinfo {author} {\bibfnamefont {X.~L.}\ \bibnamefont
  {Su}}\ and\ \bibinfo {author} {\bibfnamefont {C.~D.}\ \bibnamefont {Lin}},\
  }\bibfield  {title} {\enquote {\bibinfo {title} {Unsteady detonation with
  thermodynamic nonequilibrium effect based on the kinetic theory},}\ }\href
  {\doibase {}} {\bibfield  {journal} {\bibinfo  {journal} {Commun. Theor.
  Phys.}\ }\textbf {\bibinfo {volume} {75}},\ \bibinfo {pages} {075601}
  (\bibinfo {year} {2023})}\BibitemShut {NoStop}%
\bibitem [{\citenamefont {Lin}, \citenamefont {Luo},\ and\ \citenamefont
  {Lai}(2024)}]{lin2024discrete}%
  \BibitemOpen
  \bibfield  {author} {\bibinfo {author} {\bibfnamefont {C.~D.}\ \bibnamefont
  {Lin}}, \bibinfo {author} {\bibfnamefont {K.~H.}\ \bibnamefont {Luo}}, \ and\
  \bibinfo {author} {\bibfnamefont {H.~L.}\ \bibnamefont {Lai}},\ }\bibfield
  {title} {\enquote {\bibinfo {title} {{Discrete Boltzmann model with split
  collision for nonequilibrium reactive flows}},}\ }\href {\doibase {}}
  {\bibfield  {journal} {\bibinfo  {journal} {Commun. Theor. Phys.}\ }\textbf
  {\bibinfo {volume} {76}},\ \bibinfo {pages} {085602} (\bibinfo {year}
  {2024})}\BibitemShut {NoStop}%
\bibitem [{\citenamefont {Gan}\ \emph {et~al.}(2015)\citenamefont {Gan},
  \citenamefont {Xu}, \citenamefont {Zhang},\ and\ \citenamefont
  {Succi}}]{gan2015discrete}%
  \BibitemOpen
  \bibfield  {author} {\bibinfo {author} {\bibfnamefont {Y.~B.}\ \bibnamefont
  {Gan}}, \bibinfo {author} {\bibfnamefont {A.~G.}\ \bibnamefont {Xu}},
  \bibinfo {author} {\bibfnamefont {G.~C.}\ \bibnamefont {Zhang}}, \ and\
  \bibinfo {author} {\bibfnamefont {S.}~\bibnamefont {Succi}},\ }\bibfield
  {title} {\enquote {\bibinfo {title} {{Discrete Boltzmann modeling of
  multiphase flows: Hydrodynamic and thermodynamic non-equilibrium effects}},}\
  }\href {\doibase {}} {\bibfield  {journal} {\bibinfo  {journal} {Soft
  Matter}\ }\textbf {\bibinfo {volume} {11}},\ \bibinfo {pages} {5336--5345}
  (\bibinfo {year} {2015})}\BibitemShut {NoStop}%
\bibitem [{\citenamefont {Zhang}\ \emph {et~al.}(2019)\citenamefont {Zhang},
  \citenamefont {Xu}, \citenamefont {Zhang}, \citenamefont {Gan}, \citenamefont
  {Chen},\ and\ \citenamefont {Succi}}]{zhang2019entropy}%
  \BibitemOpen
  \bibfield  {author} {\bibinfo {author} {\bibfnamefont {Y.~D.}\ \bibnamefont
  {Zhang}}, \bibinfo {author} {\bibfnamefont {A.~G.}\ \bibnamefont {Xu}},
  \bibinfo {author} {\bibfnamefont {G.~C.}\ \bibnamefont {Zhang}}, \bibinfo
  {author} {\bibfnamefont {Y.~B.}\ \bibnamefont {Gan}}, \bibinfo {author}
  {\bibfnamefont {Z.~H.}\ \bibnamefont {Chen}}, \ and\ \bibinfo {author}
  {\bibfnamefont {S.}~\bibnamefont {Succi}},\ }\bibfield  {title} {\enquote
  {\bibinfo {title} {Entropy production in thermal phase separation: a
  kinetic-theory approach},}\ }\href {\doibase {}} {\bibfield  {journal}
  {\bibinfo  {journal} {Soft Matter}\ }\textbf {\bibinfo {volume} {15}},\
  \bibinfo {pages} {2245--2259} (\bibinfo {year} {2019})}\BibitemShut {NoStop}%
\bibitem [{\citenamefont {Gan}\ \emph {et~al.}(2022)\citenamefont {Gan},
  \citenamefont {Xu}, \citenamefont {Lai}, \citenamefont {Li}, \citenamefont
  {Sun},\ and\ \citenamefont {Succi}}]{gan2022discrete}%
  \BibitemOpen
  \bibfield  {author} {\bibinfo {author} {\bibfnamefont {Y.~B.}\ \bibnamefont
  {Gan}}, \bibinfo {author} {\bibfnamefont {A.~G.}\ \bibnamefont {Xu}},
  \bibinfo {author} {\bibfnamefont {H.~L.}\ \bibnamefont {Lai}}, \bibinfo
  {author} {\bibfnamefont {W.}~\bibnamefont {Li}}, \bibinfo {author}
  {\bibfnamefont {G.~L.}\ \bibnamefont {Sun}}, \ and\ \bibinfo {author}
  {\bibfnamefont {S.}~\bibnamefont {Succi}},\ }\bibfield  {title} {\enquote
  {\bibinfo {title} {Discrete boltzmann multi-scale modelling of
  non-equilibrium multiphase flows},}\ }\href {\doibase {}} {\bibfield
  {journal} {\bibinfo  {journal} {J. Fluid Mech.}\ }\textbf {\bibinfo {volume}
  {951}},\ \bibinfo {pages} {A8} (\bibinfo {year} {2022})}\BibitemShut
  {NoStop}%
\bibitem [{\citenamefont {Sun}\ \emph {et~al.}(2024)\citenamefont {Sun},
  \citenamefont {Gan}, \citenamefont {Xu},\ and\ \citenamefont
  {Shi}}]{sun2024droplet}%
  \BibitemOpen
  \bibfield  {author} {\bibinfo {author} {\bibfnamefont {G.~L.}\ \bibnamefont
  {Sun}}, \bibinfo {author} {\bibfnamefont {Y.~B.}\ \bibnamefont {Gan}},
  \bibinfo {author} {\bibfnamefont {A.~G.}\ \bibnamefont {Xu}}, \ and\ \bibinfo
  {author} {\bibfnamefont {Q.~F.}\ \bibnamefont {Shi}},\ }\bibfield  {title}
  {\enquote {\bibinfo {title} {{Droplet coalescence kinetics: Thermodynamic
  non-equilibrium effects and entropy production mechanism}},}\ }\href
  {\doibase {}} {\bibfield  {journal} {\bibinfo  {journal} {Phys. Fluids}\
  }\textbf {\bibinfo {volume} {36}} (\bibinfo {year} {2024}),\ {}}\BibitemShut
  {NoStop}%
\bibitem [{\citenamefont {Ye}\ \emph {et~al.}(2020)\citenamefont {Ye},
  \citenamefont {Lai}, \citenamefont {Li}, \citenamefont {Gan}, \citenamefont
  {Lin}, \citenamefont {Chen},\ and\ \citenamefont {Xu}}]{ye2020knudsen}%
  \BibitemOpen
  \bibfield  {author} {\bibinfo {author} {\bibfnamefont {H.~Y.}\ \bibnamefont
  {Ye}}, \bibinfo {author} {\bibfnamefont {H.~L.}\ \bibnamefont {Lai}},
  \bibinfo {author} {\bibfnamefont {D.~M.}\ \bibnamefont {Li}}, \bibinfo
  {author} {\bibfnamefont {Y.~B.}\ \bibnamefont {Gan}}, \bibinfo {author}
  {\bibfnamefont {C.~D.}\ \bibnamefont {Lin}}, \bibinfo {author} {\bibfnamefont
  {L.}~\bibnamefont {Chen}}, \ and\ \bibinfo {author} {\bibfnamefont {A.~G.}\
  \bibnamefont {Xu}},\ }\bibfield  {title} {\enquote {\bibinfo {title}
  {{Knudsen number effects on two-dimensional Rayleigh--Taylor instability in
  compressible fluid: Based on a discrete Boltzmann method}},}\ }\href
  {\doibase {}} {\bibfield  {journal} {\bibinfo  {journal} {Entropy}\ }\textbf
  {\bibinfo {volume} {22}},\ \bibinfo {pages} {500} (\bibinfo {year}
  {2020})}\BibitemShut {NoStop}%
\bibitem [{\citenamefont {Chen}\ \emph {et~al.}(2021)\citenamefont {Chen},
  \citenamefont {Lai}, \citenamefont {Lin},\ and\ \citenamefont
  {Li}}]{chen2021specific}%
  \BibitemOpen
  \bibfield  {author} {\bibinfo {author} {\bibfnamefont {L.}~\bibnamefont
  {Chen}}, \bibinfo {author} {\bibfnamefont {H.~L.}\ \bibnamefont {Lai}},
  \bibinfo {author} {\bibfnamefont {C.~D.}\ \bibnamefont {Lin}}, \ and\
  \bibinfo {author} {\bibfnamefont {D.~M.}\ \bibnamefont {Li}},\ }\bibfield
  {title} {\enquote {\bibinfo {title} {{Specific heat ratio effects of
  compressible Rayleigh--Taylor instability studied by discrete Boltzmann
  method}},}\ }\href {\doibase {}} {\bibfield  {journal} {\bibinfo  {journal}
  {Front. Phys.}\ }\textbf {\bibinfo {volume} {16}},\ \bibinfo {pages} {52500}
  (\bibinfo {year} {2021})}\BibitemShut {NoStop}%
\bibitem [{\citenamefont {Li}\ \emph {et~al.}(2022{\natexlab{a}})\citenamefont
  {Li}, \citenamefont {Xu}, \citenamefont {Zhang},\ and\ \citenamefont
  {Shan}}]{li2022rayleigh}%
  \BibitemOpen
  \bibfield  {author} {\bibinfo {author} {\bibfnamefont {H.~W.}\ \bibnamefont
  {Li}}, \bibinfo {author} {\bibfnamefont {A.~G.}\ \bibnamefont {Xu}}, \bibinfo
  {author} {\bibfnamefont {G.}~\bibnamefont {Zhang}}, \ and\ \bibinfo {author}
  {\bibfnamefont {Y.~M.}\ \bibnamefont {Shan}},\ }\bibfield  {title} {\enquote
  {\bibinfo {title} {{Rayleigh--Taylor instability under multi-mode
  perturbation: Discrete Boltzmann modeling with tracers}},}\ }\href {\doibase
  {}} {\bibfield  {journal} {\bibinfo  {journal} {Commun. Theor. Phys.}\
  }\textbf {\bibinfo {volume} {74}},\ \bibinfo {pages} {115601} (\bibinfo
  {year} {2022}{\natexlab{a}})}\BibitemShut {NoStop}%
\bibitem [{\citenamefont {Lai}\ \emph {et~al.}(2024)\citenamefont {Lai},
  \citenamefont {Li}, \citenamefont {Lin}, \citenamefont {Chen}, \citenamefont
  {Ye},\ and\ \citenamefont {Zhu}}]{lai2024investigation}%
  \BibitemOpen
  \bibfield  {author} {\bibinfo {author} {\bibfnamefont {H.~L.}\ \bibnamefont
  {Lai}}, \bibinfo {author} {\bibfnamefont {D.~M.}\ \bibnamefont {Li}},
  \bibinfo {author} {\bibfnamefont {C.~D.}\ \bibnamefont {Lin}}, \bibinfo
  {author} {\bibfnamefont {L.}~\bibnamefont {Chen}}, \bibinfo {author}
  {\bibfnamefont {H.~Y.}\ \bibnamefont {Ye}}, \ and\ \bibinfo {author}
  {\bibfnamefont {J.~J.}\ \bibnamefont {Zhu}},\ }\bibfield  {title} {\enquote
  {\bibinfo {title} {{Investigation of effects of initial interface conditions
  on the two-dimensional single-mode compressible Rayleigh--Taylor instability:
  Based on the discrete Boltzmann method}},}\ }\href {\doibase {}} {\bibfield
  {journal} {\bibinfo  {journal} {Comput. Fluids}\ }\textbf {\bibinfo {volume}
  {277}},\ \bibinfo {pages} {106289} (\bibinfo {year} {2024})}\BibitemShut
  {NoStop}%
\bibitem [{\citenamefont {Chen}\ \emph
  {et~al.}(2024{\natexlab{b}})\citenamefont {Chen}, \citenamefont {Xu},
  \citenamefont {Song}, \citenamefont {Gan}, \citenamefont {Zhang},\ and\
  \citenamefont {Guan}}]{chen2024surface}%
  \BibitemOpen
  \bibfield  {author} {\bibinfo {author} {\bibfnamefont {F.}~\bibnamefont
  {Chen}}, \bibinfo {author} {\bibfnamefont {A.~G.}\ \bibnamefont {Xu}},
  \bibinfo {author} {\bibfnamefont {J.~H.}\ \bibnamefont {Song}}, \bibinfo
  {author} {\bibfnamefont {Y.~B.}\ \bibnamefont {Gan}}, \bibinfo {author}
  {\bibfnamefont {Y.~D.}\ \bibnamefont {Zhang}}, \ and\ \bibinfo {author}
  {\bibfnamefont {N.}~\bibnamefont {Guan}},\ }\bibfield  {title} {\enquote
  {\bibinfo {title} {{Surface tension effects on Rayleigh--Taylor instability
  in nonideal fluids: A multiple-relaxation-time discrete Boltzmann study}},}\
  }\href {\doibase {}} {\bibfield  {journal} {\bibinfo  {journal} {Sci. China
  Phys. Mech. Astron.}\ }\textbf {\bibinfo {volume} {67}},\ \bibinfo {pages}
  {124611} (\bibinfo {year} {2024}{\natexlab{b}})}\BibitemShut {NoStop}%
\bibitem [{\citenamefont {Gan}\ \emph {et~al.}(2019)\citenamefont {Gan},
  \citenamefont {Xu}, \citenamefont {Zhang}, \citenamefont {Lin}, \citenamefont
  {Lai},\ and\ \citenamefont {Liu}}]{gan2019nonequilibrium}%
  \BibitemOpen
  \bibfield  {author} {\bibinfo {author} {\bibfnamefont {Y.~B.}\ \bibnamefont
  {Gan}}, \bibinfo {author} {\bibfnamefont {A.~G.}\ \bibnamefont {Xu}},
  \bibinfo {author} {\bibfnamefont {G.~C.}\ \bibnamefont {Zhang}}, \bibinfo
  {author} {\bibfnamefont {C.~D.}\ \bibnamefont {Lin}}, \bibinfo {author}
  {\bibfnamefont {H.~L.}\ \bibnamefont {Lai}}, \ and\ \bibinfo {author}
  {\bibfnamefont {Z.~P.}\ \bibnamefont {Liu}},\ }\bibfield  {title} {\enquote
  {\bibinfo {title} {{Nonequilibrium and morphological characterizations of
  Kelvin--Helmholtz instability in compressible flows}},}\ }\href {\doibase {}}
  {\bibfield  {journal} {\bibinfo  {journal} {Front. Phys.}\ }\textbf {\bibinfo
  {volume} {14}},\ \bibinfo {pages} {1--17} (\bibinfo {year}
  {2019})}\BibitemShut {NoStop}%
\bibitem [{\citenamefont {Li}\ \emph {et~al.}(2022{\natexlab{b}})\citenamefont
  {Li}, \citenamefont {Lai}, \citenamefont {Lin},\ and\ \citenamefont
  {Li}}]{li2022influence}%
  \BibitemOpen
  \bibfield  {author} {\bibinfo {author} {\bibfnamefont {Y.~F.}\ \bibnamefont
  {Li}}, \bibinfo {author} {\bibfnamefont {H.~L.}\ \bibnamefont {Lai}},
  \bibinfo {author} {\bibfnamefont {C.~D.}\ \bibnamefont {Lin}}, \ and\
  \bibinfo {author} {\bibfnamefont {D.~M.}\ \bibnamefont {Li}},\ }\bibfield
  {title} {\enquote {\bibinfo {title} {{Influence of the tangential velocity on
  the compressible Kelvin-Helmholtz instability with nonequilibrium
  effects}},}\ }\href {\doibase {}} {\bibfield  {journal} {\bibinfo  {journal}
  {Front. Phys}\ }\textbf {\bibinfo {volume} {17}},\ \bibinfo {pages} {63500}
  (\bibinfo {year} {2022}{\natexlab{b}})}\BibitemShut {NoStop}%
\bibitem [{\citenamefont {Li}\ and\ \citenamefont {Lin}(2024)}]{li2024kinetic}%
  \BibitemOpen
  \bibfield  {author} {\bibinfo {author} {\bibfnamefont {Y.~F.}\ \bibnamefont
  {Li}}\ and\ \bibinfo {author} {\bibfnamefont {C.~D.}\ \bibnamefont {Lin}},\
  }\bibfield  {title} {\enquote {\bibinfo {title} {{Kinetic investigation of
  Kelvin--Helmholtz instability with nonequilibrium effects in a force
  field}},}\ }\href {\doibase {}} {\bibfield  {journal} {\bibinfo  {journal}
  {Phys. Fluids}\ }\textbf {\bibinfo {volume} {36}},\ \bibinfo {pages} {116140}
  (\bibinfo {year} {2024})}\BibitemShut {NoStop}%
\bibitem [{\citenamefont {Xu}, \citenamefont {Lin},\ and\ \citenamefont
  {Lai}(2025)}]{xu2025influence}%
  \BibitemOpen
  \bibfield  {author} {\bibinfo {author} {\bibfnamefont {H.}~\bibnamefont
  {Xu}}, \bibinfo {author} {\bibfnamefont {C.~D.}\ \bibnamefont {Lin}}, \ and\
  \bibinfo {author} {\bibfnamefont {H.~L.}\ \bibnamefont {Lai}},\ }\bibfield
  {title} {\enquote {\bibinfo {title} {{Influence of phase difference and
  amplitude ratio on Kelvin--Helmholtz instability with dual-mode interface
  perturbations}},}\ }\href {\doibase {}} {\bibfield  {journal} {\bibinfo
  {journal} {Phys. Fluids}\ }\textbf {\bibinfo {volume} {37}},\ \bibinfo
  {pages} {016132} (\bibinfo {year} {2025})}\BibitemShut {NoStop}%
\bibitem [{\citenamefont {Osborn}\ \emph {et~al.}(1995)\citenamefont {Osborn},
  \citenamefont {Orlandini}, \citenamefont {Swift}, \citenamefont {Yeomans},\
  and\ \citenamefont {Banavar}}]{osborn1995lattice}%
  \BibitemOpen
  \bibfield  {author} {\bibinfo {author} {\bibfnamefont {W.~R.}\ \bibnamefont
  {Osborn}}, \bibinfo {author} {\bibfnamefont {E.}~\bibnamefont {Orlandini}},
  \bibinfo {author} {\bibfnamefont {M.~R.}\ \bibnamefont {Swift}}, \bibinfo
  {author} {\bibfnamefont {J.~M.}\ \bibnamefont {Yeomans}}, \ and\ \bibinfo
  {author} {\bibfnamefont {J.~R.}\ \bibnamefont {Banavar}},\ }\bibfield
  {title} {\enquote {\bibinfo {title} {{Lattice Boltzmann study of hydrodynamic
  spinodal decomposition}},}\ }\href {\doibase {}} {\bibfield  {journal}
  {\bibinfo  {journal} {Phys. Rev. Lett.}\ }\textbf {\bibinfo {volume} {75}},\
  \bibinfo {pages} {4031} (\bibinfo {year} {1995})}\BibitemShut {NoStop}%
\bibitem [{\citenamefont {Swift}, \citenamefont {Osborn},\ and\ \citenamefont
  {Yeomans}(1995)}]{swift1995lattice}%
  \BibitemOpen
  \bibfield  {author} {\bibinfo {author} {\bibfnamefont {M.~R.}\ \bibnamefont
  {Swift}}, \bibinfo {author} {\bibfnamefont {W.~R.}\ \bibnamefont {Osborn}}, \
  and\ \bibinfo {author} {\bibfnamefont {J.~M.}\ \bibnamefont {Yeomans}},\
  }\bibfield  {title} {\enquote {\bibinfo {title} {{Lattice Boltzmann
  simulation of nonideal fluids}},}\ }\href {\doibase {}} {\bibfield  {journal}
  {\bibinfo  {journal} {Phys. Rev. Lett.}\ }\textbf {\bibinfo {volume} {75}},\
  \bibinfo {pages} {830} (\bibinfo {year} {1995})}\BibitemShut {NoStop}%
\bibitem [{\citenamefont {Liang}\ \emph {et~al.}(2014)\citenamefont {Liang},
  \citenamefont {Shi}, \citenamefont {Guo},\ and\ \citenamefont
  {Chai}}]{liang2014phase}%
  \BibitemOpen
  \bibfield  {author} {\bibinfo {author} {\bibfnamefont {H.}~\bibnamefont
  {Liang}}, \bibinfo {author} {\bibfnamefont {B.~C.}\ \bibnamefont {Shi}},
  \bibinfo {author} {\bibfnamefont {Z.~L.}\ \bibnamefont {Guo}}, \ and\
  \bibinfo {author} {\bibfnamefont {Z.~H.}\ \bibnamefont {Chai}},\ }\bibfield
  {title} {\enquote {\bibinfo {title} {{Phase-field-based
  multiple-relaxation-time lattice Boltzmann model for incompressible
  multiphase flows}},}\ }\href {\doibase {}} {\bibfield  {journal} {\bibinfo
  {journal} {Phys. Rev. E}\ }\textbf {\bibinfo {volume} {89}},\ \bibinfo
  {pages} {053320} (\bibinfo {year} {2014})}\BibitemShut {NoStop}%
\bibitem [{\citenamefont {Liang}\ \emph {et~al.}(2016)\citenamefont {Liang},
  \citenamefont {Li}, \citenamefont {Shi},\ and\ \citenamefont
  {Chai}}]{liang2016lattice}%
  \BibitemOpen
  \bibfield  {author} {\bibinfo {author} {\bibfnamefont {H.}~\bibnamefont
  {Liang}}, \bibinfo {author} {\bibfnamefont {Q.~X.}\ \bibnamefont {Li}},
  \bibinfo {author} {\bibfnamefont {B.~C.}\ \bibnamefont {Shi}}, \ and\
  \bibinfo {author} {\bibfnamefont {Z.~H.}\ \bibnamefont {Chai}},\ }\bibfield
  {title} {\enquote {\bibinfo {title} {{Lattice Boltzmann simulation of
  three-dimensional Rayleigh-Taylor instability}},}\ }\href {\doibase {}}
  {\bibfield  {journal} {\bibinfo  {journal} {Phys. Rev. E}\ }\textbf {\bibinfo
  {volume} {93}},\ \bibinfo {pages} {033113} (\bibinfo {year}
  {2016})}\BibitemShut {NoStop}%
\bibitem [{\citenamefont {Silva}(2025)}]{silva2025analysis}%
  \BibitemOpen
  \bibfield  {author} {\bibinfo {author} {\bibfnamefont {G.}~\bibnamefont
  {Silva}},\ }\bibfield  {title} {\enquote {\bibinfo {title} {{Analysis of the
  central-moments-based lattice Boltzmann method for the numerical modeling of
  the one-dimensional advection-diffusion equation: Equivalent finite
  difference and partial differential equations}},}\ }\href {\doibase {}}
  {\bibfield  {journal} {\bibinfo  {journal} {Comput. Fluids}\ ,\ \bibinfo
  {pages} {106535}} (\bibinfo {year} {2025})}\BibitemShut {NoStop}%
\bibitem [{\citenamefont {Xu}\ \emph {et~al.}(2012)\citenamefont {Xu},
  \citenamefont {Zhang}, \citenamefont {Gan}, \citenamefont {Chen},\ and\
  \citenamefont {Yu}}]{xu2012lattice}%
  \BibitemOpen
  \bibfield  {author} {\bibinfo {author} {\bibfnamefont {A.~G.}\ \bibnamefont
  {Xu}}, \bibinfo {author} {\bibfnamefont {G.~C.}\ \bibnamefont {Zhang}},
  \bibinfo {author} {\bibfnamefont {Y.~B.}\ \bibnamefont {Gan}}, \bibinfo
  {author} {\bibfnamefont {F.}~\bibnamefont {Chen}}, \ and\ \bibinfo {author}
  {\bibfnamefont {X.~J.}\ \bibnamefont {Yu}},\ }\bibfield  {title} {\enquote
  {\bibinfo {title} {{Lattice Boltzmann modeling and simulation of compressible
  flows}},}\ }\href {\doibase {}} {\bibfield  {journal} {\bibinfo  {journal}
  {Front. Phys}\ }\textbf {\bibinfo {volume} {7}},\ \bibinfo {pages} {582--600}
  (\bibinfo {year} {2012})}\BibitemShut {NoStop}%
\bibitem [{\citenamefont {Lin}\ \emph {et~al.}(2014)\citenamefont {Lin},
  \citenamefont {Xu}, \citenamefont {Zhang}, \citenamefont {Li},\ and\
  \citenamefont {Succi}}]{lin2014polar}%
  \BibitemOpen
  \bibfield  {author} {\bibinfo {author} {\bibfnamefont {C.~D.}\ \bibnamefont
  {Lin}}, \bibinfo {author} {\bibfnamefont {A.~G.}\ \bibnamefont {Xu}},
  \bibinfo {author} {\bibfnamefont {G.~C.}\ \bibnamefont {Zhang}}, \bibinfo
  {author} {\bibfnamefont {Y.~J.}\ \bibnamefont {Li}}, \ and\ \bibinfo {author}
  {\bibfnamefont {S.}~\bibnamefont {Succi}},\ }\bibfield  {title} {\enquote
  {\bibinfo {title} {{Polar-coordinate lattice Boltzmann modeling of
  compressible flows}},}\ }\href {\doibase {}} {\bibfield  {journal} {\bibinfo
  {journal} {Phys. Rev. E}\ }\textbf {\bibinfo {volume} {89}},\ \bibinfo
  {pages} {013307} (\bibinfo {year} {2014})}\BibitemShut {NoStop}%
\bibitem [{\citenamefont {Lin}\ \emph {et~al.}(2016)\citenamefont {Lin},
  \citenamefont {Xu}, \citenamefont {Zhang},\ and\ \citenamefont
  {Li}}]{lin2016double}%
  \BibitemOpen
  \bibfield  {author} {\bibinfo {author} {\bibfnamefont {C.~D.}\ \bibnamefont
  {Lin}}, \bibinfo {author} {\bibfnamefont {A.~G.}\ \bibnamefont {Xu}},
  \bibinfo {author} {\bibfnamefont {G.~C.}\ \bibnamefont {Zhang}}, \ and\
  \bibinfo {author} {\bibfnamefont {Y.~J.}\ \bibnamefont {Li}},\ }\bibfield
  {title} {\enquote {\bibinfo {title} {{Double-distribution-function discrete
  Boltzmann model for combustion}},}\ }\href {\doibase {}} {\bibfield
  {journal} {\bibinfo  {journal} {Combust. Flame}\ }\textbf {\bibinfo {volume}
  {164}},\ \bibinfo {pages} {137--151} (\bibinfo {year} {2016})}\BibitemShut
  {NoStop}%
\bibitem [{\citenamefont {Chen}, \citenamefont {Xu},\ and\ \citenamefont
  {Zhang}(2018)}]{chen2018collaboration}%
  \BibitemOpen
  \bibfield  {author} {\bibinfo {author} {\bibfnamefont {F.}~\bibnamefont
  {Chen}}, \bibinfo {author} {\bibfnamefont {A.~G.}\ \bibnamefont {Xu}}, \ and\
  \bibinfo {author} {\bibfnamefont {G.~C.}\ \bibnamefont {Zhang}},\ }\bibfield
  {title} {\enquote {\bibinfo {title} {{Collaboration and competition between
  Richtmyer--Meshkov instability and Rayleigh--Taylor instability}},}\ }\href
  {\doibase {}} {\bibfield  {journal} {\bibinfo  {journal} {Phys. Fluids}\
  }\textbf {\bibinfo {volume} {30}},\ \bibinfo {pages} {102105} (\bibinfo
  {year} {2018})}\BibitemShut {NoStop}%
\bibitem [{\citenamefont {Shan}\ \emph {et~al.}(2023)\citenamefont {Shan},
  \citenamefont {Xu}, \citenamefont {Wang},\ and\ \citenamefont
  {Zhang}}]{shan2023nonequilibrium}%
  \BibitemOpen
  \bibfield  {author} {\bibinfo {author} {\bibfnamefont {Y.~M.}\ \bibnamefont
  {Shan}}, \bibinfo {author} {\bibfnamefont {A.~G.}\ \bibnamefont {Xu}},
  \bibinfo {author} {\bibfnamefont {L.~F.}\ \bibnamefont {Wang}}, \ and\
  \bibinfo {author} {\bibfnamefont {Y.~D.}\ \bibnamefont {Zhang}},\ }\bibfield
  {title} {\enquote {\bibinfo {title} {{Nonequilibrium kinetics effects in
  Richtmyer--Meshkov instability and reshock processes}},}\ }\href {\doibase
  {}} {\bibfield  {journal} {\bibinfo  {journal} {Commun. Theor. Phys.}\
  }\textbf {\bibinfo {volume} {75}},\ \bibinfo {pages} {115601} (\bibinfo
  {year} {2023})}\BibitemShut {NoStop}%
\bibitem [{\citenamefont {Yang}\ \emph {et~al.}(2023)\citenamefont {Yang},
  \citenamefont {Lin}, \citenamefont {Li},\ and\ \citenamefont
  {Lai}}]{yang2023influence}%
  \BibitemOpen
  \bibfield  {author} {\bibinfo {author} {\bibfnamefont {T.}~\bibnamefont
  {Yang}}, \bibinfo {author} {\bibfnamefont {C.~D.}\ \bibnamefont {Lin}},
  \bibinfo {author} {\bibfnamefont {D.~M.}\ \bibnamefont {Li}}, \ and\ \bibinfo
  {author} {\bibfnamefont {H.~L.}\ \bibnamefont {Lai}},\ }\bibfield  {title}
  {\enquote {\bibinfo {title} {{Influence of density ratios on
  Richtmyer--Meshkov instability with non-equilibrium effects in the reshock
  process}},}\ }\href {\doibase {}} {\bibfield  {journal} {\bibinfo  {journal}
  {Inventions}\ }\textbf {\bibinfo {volume} {8}},\ \bibinfo {pages} {157}
  (\bibinfo {year} {2023})}\BibitemShut {NoStop}%
\bibitem [{\citenamefont {Song}\ \emph {et~al.}(2024)\citenamefont {Song},
  \citenamefont {Xu}, \citenamefont {Miao}, \citenamefont {Chen}, \citenamefont
  {Liu}, \citenamefont {Wang}, \citenamefont {Wang},\ and\ \citenamefont
  {Hou}}]{song2024plasma}%
  \BibitemOpen
  \bibfield  {author} {\bibinfo {author} {\bibfnamefont {J.~H.}\ \bibnamefont
  {Song}}, \bibinfo {author} {\bibfnamefont {A.~G.}\ \bibnamefont {Xu}},
  \bibinfo {author} {\bibfnamefont {L.}~\bibnamefont {Miao}}, \bibinfo {author}
  {\bibfnamefont {F.}~\bibnamefont {Chen}}, \bibinfo {author} {\bibfnamefont
  {Z.~P.}\ \bibnamefont {Liu}}, \bibinfo {author} {\bibfnamefont {L.~F.}\
  \bibnamefont {Wang}}, \bibinfo {author} {\bibfnamefont {N.~F.}\ \bibnamefont
  {Wang}}, \ and\ \bibinfo {author} {\bibfnamefont {X.}~\bibnamefont {Hou}},\
  }\bibfield  {title} {\enquote {\bibinfo {title} {{Plasma kinetics: Discrete
  Boltzmann modeling and Richtmyer--Meshkov instability}},}\ }\href {\doibase
  {}} {\bibfield  {journal} {\bibinfo  {journal} {Phys. Fluids}\ }\textbf
  {\bibinfo {volume} {36}},\ \bibinfo {pages} {016107} (\bibinfo {year}
  {2024})}\BibitemShut {NoStop}%
\bibitem [{\citenamefont {Weber}\ \emph {et~al.}(2014)\citenamefont {Weber},
  \citenamefont {Clark}, \citenamefont {Cook}, \citenamefont {Busby},\ and\
  \citenamefont {Robey}}]{weber2014inhibition}%
  \BibitemOpen
  \bibfield  {author} {\bibinfo {author} {\bibfnamefont {C.~R.}\ \bibnamefont
  {Weber}}, \bibinfo {author} {\bibfnamefont {D.~S.}\ \bibnamefont {Clark}},
  \bibinfo {author} {\bibfnamefont {A.~W.}\ \bibnamefont {Cook}}, \bibinfo
  {author} {\bibfnamefont {L.~E.}\ \bibnamefont {Busby}}, \ and\ \bibinfo
  {author} {\bibfnamefont {H.~F.}\ \bibnamefont {Robey}},\ }\bibfield  {title}
  {\enquote {\bibinfo {title} {Inhibition of turbulence in
  inertial-confinement-fusion hot spots by viscous dissipation},}\ }\href
  {\doibase {}} {\bibfield  {journal} {\bibinfo  {journal} {Phys. Rev. E}\
  }\textbf {\bibinfo {volume} {89}},\ \bibinfo {pages} {053106} (\bibinfo
  {year} {2014})}\BibitemShut {NoStop}%
\bibitem [{\citenamefont {Pereira}\ \emph {et~al.}(2021)\citenamefont
  {Pereira}, \citenamefont {Grinstein}, \citenamefont {Israel},\ and\
  \citenamefont {Rauenzahn}}]{pereira2021molecular}%
  \BibitemOpen
  \bibfield  {author} {\bibinfo {author} {\bibfnamefont {F.~S.}\ \bibnamefont
  {Pereira}}, \bibinfo {author} {\bibfnamefont {F.~F.}\ \bibnamefont
  {Grinstein}}, \bibinfo {author} {\bibfnamefont {D.~M.}\ \bibnamefont
  {Israel}}, \ and\ \bibinfo {author} {\bibfnamefont {R.}~\bibnamefont
  {Rauenzahn}},\ }\bibfield  {title} {\enquote {\bibinfo {title} {Molecular
  viscosity and diffusivity effects in transitional and shock-driven mixing
  flows},}\ }\href {\doibase {}} {\bibfield  {journal} {\bibinfo  {journal}
  {Phys. Rev. E}\ }\textbf {\bibinfo {volume} {103}},\ \bibinfo {pages}
  {013106} (\bibinfo {year} {2021})}\BibitemShut {NoStop}%
\bibitem [{\citenamefont {Mikaelian}(1993)}]{mikaelian1993effect}%
  \BibitemOpen
  \bibfield  {author} {\bibinfo {author} {\bibfnamefont {K.~O.}\ \bibnamefont
  {Mikaelian}},\ }\bibfield  {title} {\enquote {\bibinfo {title} {{Effect of
  viscosity on Rayleigh--Taylor and Richtmyer--Meshkov instabilities}},}\
  }\href {\doibase {}} {\bibfield  {journal} {\bibinfo  {journal} {Phys. Rev.
  E}\ }\textbf {\bibinfo {volume} {47}},\ \bibinfo {pages} {375} (\bibinfo
  {year} {1993})}\BibitemShut {NoStop}%
\bibitem [{\citenamefont {Sohn}(2009)}]{sohn2009effects}%
  \BibitemOpen
  \bibfield  {author} {\bibinfo {author} {\bibfnamefont {S.~I.}\ \bibnamefont
  {Sohn}},\ }\bibfield  {title} {\enquote {\bibinfo {title} {{Effects of
  surface tension and viscosity on the growth rates of Rayleigh--Taylor and
  Richtmyer--Meshkov instabilities}},}\ }\href {\doibase {}} {\bibfield
  {journal} {\bibinfo  {journal} {Phys. Rev. E}\ }\textbf {\bibinfo {volume}
  {80}},\ \bibinfo {pages} {055302} (\bibinfo {year} {2009})}\BibitemShut
  {NoStop}%
\bibitem [{\citenamefont {Walchli}\ and\ \citenamefont
  {Thornber}(2017)}]{walchli2017reynolds}%
  \BibitemOpen
  \bibfield  {author} {\bibinfo {author} {\bibfnamefont {B.}~\bibnamefont
  {Walchli}}\ and\ \bibinfo {author} {\bibfnamefont {B.}~\bibnamefont
  {Thornber}},\ }\bibfield  {title} {\enquote {\bibinfo {title} {{Reynolds
  number effects on the single-mode Richtmyer--Meshkov instability}},}\ }\href
  {\doibase {}} {\bibfield  {journal} {\bibinfo  {journal} {Phys. Rev. E}\
  }\textbf {\bibinfo {volume} {95}},\ \bibinfo {pages} {013104} (\bibinfo
  {year} {2017})}\BibitemShut {NoStop}%
\bibitem [{\citenamefont {Wong}, \citenamefont {Livescu},\ and\ \citenamefont
  {Lele}(2019)}]{wong2019high}%
  \BibitemOpen
  \bibfield  {author} {\bibinfo {author} {\bibfnamefont {M.~L.}\ \bibnamefont
  {Wong}}, \bibinfo {author} {\bibfnamefont {D.}~\bibnamefont {Livescu}}, \
  and\ \bibinfo {author} {\bibfnamefont {S.~K.}\ \bibnamefont {Lele}},\
  }\bibfield  {title} {\enquote {\bibinfo {title} {{High-resolution
  Navier--Stokes simulations of Richtmyer--Meshkov instability with
  reshock}},}\ }\href {\doibase {}} {\bibfield  {journal} {\bibinfo  {journal}
  {Phys. Rev. Fluids}\ }\textbf {\bibinfo {volume} {4}},\ \bibinfo {pages}
  {104609} (\bibinfo {year} {2019})}\BibitemShut {NoStop}%
\bibitem [{\citenamefont {Liu}\ \emph {et~al.}(2020)\citenamefont {Liu},
  \citenamefont {Yu}, \citenamefont {Chen}, \citenamefont {Zhang},
  \citenamefont {Xu},\ and\ \citenamefont {Liu}}]{liu2020contribution}%
  \BibitemOpen
  \bibfield  {author} {\bibinfo {author} {\bibfnamefont {H.~C.}\ \bibnamefont
  {Liu}}, \bibinfo {author} {\bibfnamefont {B.}~\bibnamefont {Yu}}, \bibinfo
  {author} {\bibfnamefont {H.}~\bibnamefont {Chen}}, \bibinfo {author}
  {\bibfnamefont {B.}~\bibnamefont {Zhang}}, \bibinfo {author} {\bibfnamefont
  {H.}~\bibnamefont {Xu}}, \ and\ \bibinfo {author} {\bibfnamefont
  {H.}~\bibnamefont {Liu}},\ }\bibfield  {title} {\enquote {\bibinfo {title}
  {{Contribution of viscosity to the circulation deposition in the
  Richtmyer--Meshkov instability}},}\ }\href {\doibase {}} {\bibfield
  {journal} {\bibinfo  {journal} {J. Fluid Mech.}\ }\textbf {\bibinfo {volume}
  {895}},\ \bibinfo {pages} {A10} (\bibinfo {year} {2020})}\BibitemShut
  {NoStop}%
\bibitem [{\citenamefont {Sun}, \citenamefont {Wang},\ and\ \citenamefont
  {Piriz}(2020)}]{sun2020unified}%
  \BibitemOpen
  \bibfield  {author} {\bibinfo {author} {\bibfnamefont {Y.~B.}\ \bibnamefont
  {Sun}}, \bibinfo {author} {\bibfnamefont {C.}~\bibnamefont {Wang}}, \ and\
  \bibinfo {author} {\bibfnamefont {A.~R.}\ \bibnamefont {Piriz}},\ }\bibfield
  {title} {\enquote {\bibinfo {title} {{A unified model to study the effects of
  elasticity, viscosity, and magnetic fields on linear Richtmyer--Meshkov
  instability}},}\ }\href {\doibase {}} {\bibfield  {journal} {\bibinfo
  {journal} {J. Appl. Phys.}\ }\textbf {\bibinfo {volume} {128}},\ \bibinfo
  {pages} {125901} (\bibinfo {year} {2020})}\BibitemShut {NoStop}%
\bibitem [{\citenamefont {Sofonea}\ and\ \citenamefont
  {Sekerka}(2001)}]{sofonea2001bgk}%
  \BibitemOpen
  \bibfield  {author} {\bibinfo {author} {\bibfnamefont {V.}~\bibnamefont
  {Sofonea}}\ and\ \bibinfo {author} {\bibfnamefont {R.~F.}\ \bibnamefont
  {Sekerka}},\ }\bibfield  {title} {\enquote {\bibinfo {title} {{BGK models for
  diffusion in isothermal binary fluid systems}},}\ }\href {\doibase {}}
  {\bibfield  {journal} {\bibinfo  {journal} {Physica A}\ }\textbf {\bibinfo
  {volume} {299}},\ \bibinfo {pages} {494--520} (\bibinfo {year}
  {2001})}\BibitemShut {NoStop}%
\bibitem [{\citenamefont {Watari}\ and\ \citenamefont
  {Tsutahara}(2004)}]{watari2004possibility}%
  \BibitemOpen
  \bibfield  {author} {\bibinfo {author} {\bibfnamefont {M.}~\bibnamefont
  {Watari}}\ and\ \bibinfo {author} {\bibfnamefont {M.}~\bibnamefont
  {Tsutahara}},\ }\bibfield  {title} {\enquote {\bibinfo {title} {{Possibility
  of constructing a multispeed Bhatnagar-Gross-Krook thermal model of the
  lattice Boltzmann method}},}\ }\href {\doibase {}} {\bibfield  {journal}
  {\bibinfo  {journal} {Phys. Rev. E}\ }\textbf {\bibinfo {volume} {70}},\
  \bibinfo {pages} {016703} (\bibinfo {year} {2004})}\BibitemShut {NoStop}%
\bibitem [{\citenamefont {Lin}\ and\ \citenamefont
  {Luo}(2019)}]{lin2019discrete}%
  \BibitemOpen
  \bibfield  {author} {\bibinfo {author} {\bibfnamefont {C.~D.}\ \bibnamefont
  {Lin}}\ and\ \bibinfo {author} {\bibfnamefont {K.~H.}\ \bibnamefont {Luo}},\
  }\bibfield  {title} {\enquote {\bibinfo {title} {{Discrete Boltzmann modeling
  of unsteady reactive flows with nonequilibrium effects}},}\ }\href {\doibase
  {}} {\bibfield  {journal} {\bibinfo  {journal} {Phys. Rev. E}\ }\textbf
  {\bibinfo {volume} {99}},\ \bibinfo {pages} {012142} (\bibinfo {year}
  {2019})}\BibitemShut {NoStop}%
\bibitem [{\citenamefont {Marco}\ and\ \citenamefont
  {Oleg}(2020)}]{latini2020comparison}%
  \BibitemOpen
  \bibfield  {author} {\bibinfo {author} {\bibfnamefont {L.}~\bibnamefont
  {Marco}}\ and\ \bibinfo {author} {\bibfnamefont {S.}~\bibnamefont {Oleg}},\
  }\bibfield  {title} {\enquote {\bibinfo {title} {{A comparison of two-and
  three-dimensional single-mode reshocked Richtmyer--Meshkov instability
  growth}},}\ }\href {\doibase {}} {\bibfield  {journal} {\bibinfo  {journal}
  {Physica D}\ }\textbf {\bibinfo {volume} {401}},\ \bibinfo {pages} {132201}
  (\bibinfo {year} {2020})}\BibitemShut {NoStop}%
\end{thebibliography}%

\end{document}